\documentclass[letter]{autart}

\usepackage{amsmath}
\usepackage{amssymb,graphicx,color}
\usepackage{enumitem}
\usepackage{cite}
\usepackage{bbm}
\usepackage{algorithm,algorithmic}
\usepackage{subfigure}
\usepackage[colorlinks=true, linkcolor=blue, citecolor=blue,
urlcolor=blue]{hyperref}
\usepackage{natbib}

\usepackage{savesym}
\savesymbol{AND}

\newtheorem{theorem}{Theorem}[section]
\newtheorem{proposition}[theorem]{Proposition}
\newtheorem{lemma}[theorem]{Lemma}

\newtheorem{remark}[theorem]{Remark}

\newcommand{\longthmtitle}[1]{\mbox{}\textit{(#1).}}

\newcommand{\real}{\ensuremath{\mathbb{R}}}
\newcommand{\realpositive}{\ensuremath{\mathbb{R}_{>0}}}
\newcommand{\realnonnegative}{\ensuremath{\mathbb{R}_{\ge 0}}}
\newcommand{\integerspositive}{{\mathbb{Z}_{>0}}}
\newcommand{\integersnonnegative}{{\mathbb{Z}}_{\ge 0}}

\newcommand{\intersect}{\ensuremath{\operatorname{\cap}}}

\DeclareMathOperator*{\argmax}{argmax}

\newcommand{\Bc}{\mathcal{B}}

\newcommand{\Dc}{\mathcal{D}}

\newcommand{\Ic}{\mathcal{I}}
\newcommand{\Jc}{\mathcal{J}}
\newcommand{\Lc}{\mathcal{L}}

\newcommand{\Nc}{\mathcal{N}}
\newcommand{\Pc}{\mathcal{P}}

\newcommand{\Tc}{\mathcal{T}}

\newcommand{\Deltil}{\tilde{\Delta}}
\newcommand{\rtil}{\tilde{r}}
\newcommand{\bbar}{\bar{b}}
\newcommand{\pbar}{\bar{p}}
\newcommand{\pibar}{\bar{\pi}}
\newcommand{\DataWithSeq}{D}
\newcommand{\Tbar}{\bar{T}}
\newcommand{\Lctil}{\tilde{\Lc}}
\newcommand{\pmax}{p^{\max}}

\newcommand{\Nzf}{\Nc_{j_0}^{j_f}}
\newcommand{\phizf}{\phi_{j_0}^{j_f}}

\newcommand{\Btil}{\tilde{B}}

\newcommand{\Enorm}[1]{\|#1\|_{2}}
\newcommand{\Infnorm}[1]{\|#1\|_{\infty}}
\newcommand{\until}[1]{\{1,\dots,{#1}\}}
\newcommand{\identity}[1]{\mathrm{I}_{#1}}
\newcommand{\card}[1]{|#1|}

\newcommand{\map}[3]{#1:#2 \rightarrow #3}

\newcommand{\floor}[1]{ \lfloor #1 \rfloor }
\newcommand{\losat}[1]{ \left[ #1 \right]_+ }
\newcommand{\Bsat}[1]{ \Big[ #1 \Big]_+ }

\newcommand{\uline}[1]{\underline{#1}}
\newcommand{\triggerPf}{h_{\operatorname{pf}}}
\newcommand{\triggerPfbar}{\bar{h}_{\operatorname{pf}}}
\newcommand{\triggerCh}{h_{\operatorname{ch}}}

\newcommand{\triggerChbar}{\bar{h}_{\operatorname{ch}}}
\newcommand{\rhofun}[2]{\rho_{#1}(#2)}

\newcommand{\Datacap}[2]{ \Dc( #1, #2 ) }
\newcommand{\Subcap}[2]{ \Dc_s( #1, #2 ) }

\newcommand{\hatDcap}[2]{ \hat{\Dc}( #1, #2 ) }
\newcommand{\hatScap}[2]{ \hat{\Dc}_s( #1, #2 ) }

\newcommand{\optxbits}[1]{ \Phi^{\tau_l}( #1 ) }
\newcommand{\packbnd}[1]{ \psi^{\tau_l}( #1 ) }

\newcommand{\oprocendsymbol}{\hbox{$\bullet$}}
\newcommand{\oprocend}{\relax\ifmmode\else\unskip\hfill\fi\oprocendsymbol}

\renewcommand{\theenumi}{(\roman{enumi})}

\newcommand{\myclearpage}{\clearpage}
\renewcommand{\myclearpage}{}

\graphicspath{{figures/}}

\begin{document}

\date{\today}

\runauthor{P. Tallapragada, M. Franceschetti, and J. Cort\'es}

\begin{frontmatter}


  \title{Event-triggered control under time-varying rates \\ and
    channel blackouts}

  \thanks{A preliminary version of this paper has been submitted to
    the 2015 Allerton Conference on Communications, Control and
    Computing as~\cite{PT-MF-JC:15-allerton}.}

  \author[ucsdmae]{Pavankumar Tallapragada} \ead{ptallapragada@ucsd.edu}
  \qquad
  \author[ucsdece]{Massimo Franceschetti} \ead{massimo@ucsd.edu}
  \qquad
  \author[ucsdmae]{Jorge Cort\'es} \ead{cortes@ucsd.edu}
  
  \address[ucsdmae]{Department of Mechanical and Aerospace
    Engineering, University of California, San Diego, CA, 92093, USA}

  \address[ucsdece]{Department of Electrical and Computer Engineering,
    University of California, San Diego, CA, 92093, USA}

  \begin{abstract}
    This paper studies event-triggered stabilization of linear
    time-invariant systems over time-varying rate-limited
    communication channels. We explicitly account for the possibility
    of channel blackouts, i.e., intervals of time when the
    communication channel is unavailable for feedback. Assuming prior
    knowledge of the channel evolution, we study the data capacity,
    which is the maximum total number of bits that could be
    communicated over a given time interval, and provide an efficient
    real-time algorithm to lower bound it for a deterministic
    time-slotted model of channel evolution. Building on these
    results, we design an event-triggering strategy that guarantees
    Zeno-free, exponential stabilization at a desired convergence rate
    even in the presence of intermittent channel blackouts. The
    contributions are the notion of channel blackouts, the effective
    event-triggered control despite their occurrence, and the analysis
    and quantification of the data capacity for a class of
    time-varying continuous-time channels. Various simulations
    illustrate the results.
  \end{abstract}


\end{frontmatter}

\section{Introduction}

Control under communication constraints has key theoretical and
practical importance given the increasing ubiquity of networked
cyber-physical systems in nearly every aspect of modern life,
including transportation, energy, agriculture, and healthcare. This
has motivated a vast amount of research to address the challenges
posed by communication channels with limited, time-varying, and
unreliable bit rates.  This paper is a contribution to the growing
body of results that employ either information-theoretic or
opportunistic triggered control to address the problem of
stabilization under constrained resources.  Specifically, we seek to
combine both approaches to deal with the control of linear
time-invariant systems under time-varying channels, including for the
possibility of blackouts, i.e., intervals of time during which the
channel is completely unavailable for control. Applications where
these channel models are useful include communication in contested
environments and scheduling shared communication resources.

\emph{Literature review:} The literature of information-theoretic
control under communication constraints focuses on identifying
necessary and sufficient conditions on the bit rates that guarantee
stabilization under various assumptions on the (often stochastically
modeled) communication channels.  Comprehensive overviews may be found
in~\citep{GNN-FF-SZ-RJE:07,MF-PM:14}. Early data rate
results~\citep{GNN-RJE:00,GNN-RJE:04,ST-SM:04} provided tight
necessary and sufficient conditions on the data rate of the encoded
feedback for asymptotic stabilization in the discrete-time
setting. Since then, the problem has been studied under increasingly
complex assumptions on the communication channels, see
e.g.,~\citep{NM-MD-NE:06,PM-MF-SD-GNN:09,PM-LC-MF:13}. In the
continuous-time setting, the problem has been studied under either
periodic sampling or aperiodic sampling with known upper and lower
bounds on the sampling period. The works~\citep{LL-JB:04,LL-JB:07} deal
with single-input systems,~\citep{CDP:05} deals with nonlinear
feedforward systems, and~\citep{DL:14} deals with switched linear
systems and characterizes the convergence rate of the finite data-rate
stabilization scheme. The recent work~\citep{JP-JPH-DL:14} explores the
stabilization problem under a state-based aperiodic transmission
policy, with the inter-transmission intervals being integral multiples
of a fixed stepsize.  In general, this literature has not explored the
potential advantages of tuning the sampling period in the periodic
case or if state-based aperiodic sampling can provide any gains in
efficiency and performance.  On the other hand, the event-triggered
approach, see e.g.~\citep{PT:07,XW-MDL:11,WPMHH-KHJ-PT:12} and
references therein, exploits the tolerance to measurement errors to
design goal-driven, opportunistic state-based aperiodic sampling. The
literature on event-triggered control mainly focuses on guaranteeing
control performance while minimizing the number of transmissions but
largely ignores quantization, data capacity, and other important
aspects of communication. Some of the few exceptions
include~\citep{PT-NC:12,EG-PJA:13}, which utilize static logarithmic
quantization and~\citep{DL-JL:10, LL-XW-MDL:12a, YS-XW:14} (see also
references therein) which use dynamic quantization. All these works
guarantee a positive lower bound on the inter-transmission times,
while~\citep{DL-JL:10, LL-XW-MDL:12a, YS-XW:14} also provide a uniform
bound on the communication bit rate (i.e., the number of bits per
transmission). However, these references do not address the inverse
problem of triggering and quantization given a limit on the
communication bit rate. Moreover, the channel is assumed to always be
available to the control system and hence event-triggered designs
typically do not take into account the possibility of channel
blackouts. An important exception to this statement
is~\citep{AA-PT:09}, which uses the deadlines generated by a
self-triggered controller to perform a kind of instantaneous or
short-term scheduling. However, if the communication latency is
time-varying either because of a time-varying channel or because of
time-varying packet sizes, which is important in finite precision
feedback control, it is difficult to guarantee long-term future
schedulability and system performance.
%
%
Our recent work~\citep{PT-JC:16-tac} combines the information-theoretic
and event-triggered control approaches to address the problem of
event-triggered stabilization of continuous-time linear time-invariant
systems under bounded bit rates. The event-triggered formulation
allows us to guarantee, in the absence of channel blackouts, a
specified rate of convergence in the presence of non-instantaneous
communication and possibly time-varying communication rate. The
incorporation of information-theoretic aspects in our design also
allows us to analyze sufficient average data rate, something usually
absent in the event-triggered literature.

\emph{Statement of contributions:} We combine information-theoretic
and event-triggered control to address the stabilization problem for
linear time-invariant systems over time-varying rate-limited
communication channels that may be subject to sporadic blackouts.  Our
starting point is a description of the communication channel through
two time-varying channel functions representing, respectively, the
minimum instantaneous communication-rate and the maximum packet size
that can be successfully transmitted. Our model explicitly accounts
for the possibility of channel blackouts, which are intervals of time
during which no packet can be successfully transmitted.  Our first
contribution is the definition of the concept of data capacity, i.e.,
the maximum number of bits that may be communicated over possibly
multiple transmissions during an arbitrary time interval under
complete knowledge of the channel evolution. This concept plays a key
role in effectively controlling the system despite the occurrence of
blackouts. The computation of data capacity for general time-varying
channels is challenging.  We show that, for the class of piecewise
constant channel functions, the computation of data capacity can be
formulated as an allocation problem involving the number of bits to be
transmitted over each interval where the channel functions are
constant. This equivalence sets the basis for our second contribution,
which is the design of an algorithm to lower bound in real time the
data capacity over an arbitrary time interval.  Our third and final
contribution is the synthesis of event-triggered control schemes that,
using prior knowledge of the channel information, plan the
transmissions in order to guarantee the exponential stabilization of
the system at a desired convergence rate, even in the presence of
intermittent channel blackouts.  Our design critically relies on three
elements: a performance-trigger function that measures how close the
system state is to violating the control objective, a channel-trigger
function that keeps track of the number of bits required at any moment
to guarantee performance at least for a certain period of time in the
future, and the lower bounds on data capacity provided by our
real-time algorithm.  Our notion of scheduled channel blackouts and
stabilization despite their occurrence is a key contribution in the
context of event-triggered control, which typically assumes the
channel is available for feedback on demand. Various simulations
illustrate our results.


\emph{Notation:} We let $\real$, $\realnonnegative$,
$\integerspositive$, and $\integersnonnegative$ denote the set of
real, nonnegative real, positive integer, and nonnegative integer
numbers, respectively.  We let $\card{S}$ denote the cardinality of
the set~$S$. We denote by $\Enorm{.}$ and $\Infnorm{.}$ the Euclidean
and infinity norm of a vector, respectively, or the corresponding
induced norm of a matrix. For a symmetric matrix $A \in \real^{n
  \times n}$, we let $\lambda_m(A)$ and $\lambda_M(A)$ denote its
smallest and largest eigenvalues, respectively. For any matrix norm
$\|.\|$, note that $\|e^{A \tau}\| \leq e^{\|A\| \tau}$. For a number
$a \in \real$, we let $\losat{a} \triangleq \max \{ 0, a \}$. For a
function $f: \real \mapsto \real^{n}$ and any $t \in \real$, we let
$f(t^-)$ and $f(t^+)$ denote the limit from the left, $\displaystyle
\lim_{s \uparrow t} f(s)$ and the limit from the right, $\displaystyle
\lim_{s \downarrow t} f(s)$, respectively.

\myclearpage
\section{Problem statement}\label{sec:problem-statement}

We start with the description of the system dynamics, then describe
the model for the communication channel, and finally state the control
objective.

\subsection{System description}

We consider a linear time-invariant control system,
\begin{align}\label{eqn:plant_dyn}
  \dot{x}(t) = A x(t) + B u(t),
\end{align}
where $x \in \real^n$ denotes the state of the plant and $u \in
\real^m$ the control input, while $A \in \real^{n \times n}$ and $B
\in \real^{n \times m}$ are the system matrices.  Our starting point
is the existence of a continuous-time feedback stabilizer of the plant
dynamics~\eqref{eqn:plant_dyn}. Formally, we select a control gain
matrix $K \in \real^{m \times n}$ such that the matrix $\bar{A} =
A+BK$ is Hurwitz. Under this assumption, the continuous-time feedback
$u(t) = K x(t)$ renders the origin of~\eqref{eqn:plant_dyn} globally
exponentially stable.

The plant is equipped with a sensor (the \emph{encoder}) and an
actuator (the \emph{decoder}) that are not co-located. The sensor can
measure the state exactly and the actuator can exert the input to the
plant with infinite precision. However, the sensor may transmit state
information to the controller at the actuator only at discrete time
instants \emph{of its choice}, using only a finite number of bits.  We
let $\{ t_k \}_{k \in \integerspositive} \subset \realnonnegative$ be
the sequence of \emph{transmission times} at which the sensor
transmits an encoded packet of data, $\{r_k\}_{k \in
  \integerspositive} \subset \realnonnegative$ the sequence of
\textit{reception times} at which the decoder receives a complete
packet of data, and $\{\rtil_k\}_{k \in \integerspositive} \subset
\realnonnegative$ the sequence of \emph{update times} at which the
decoder updates the controller state. At a transmission time $t_k$,
the sensor sends $b_k$ bits, which encode the plant state. Due to
causality, $\rtil_k \geq r_k \geq t_k$, and we denote by
\begin{equation*}
  \Delta_k \triangleq r_k - t_k, \quad \Deltil_k \triangleq \rtil_k -
  t_k ,
\end{equation*}
the $k^\text{th}$ \emph{communication time} and $k^\text{th}$
\emph{time-to-update}, respectively. The distinction between the
reception times and the update times is a generalization with respect
to our previous work~\citep{PT-JC:16-tac} and provides greater
flexibility in the presence of time-varying channels, particularly in
cases where the channel is unavailable for certain periods of time.

\subsection{Communication channel}

Our model for the time-varying communication channel is fully
determined by the map $\map{R}{\realnonnegative}{\realnonnegative}$,
where $R_a = n R$ is the \emph{minimum instantaneous
  communication-rate} at a given time, and the map
$\map{\pbar}{\realnonnegative}{\integersnonnegative}$, where $\bbar =
n \pbar$ is the \emph{maximum packet size} that can be successfully
transmitted at a given time.  More specifically, we assume the
$k^\text{th}$ communication time and the $k^\text{th}$ time-to-update
satisfy
\begin{subequations}\label{eqn:Del-cond}
  \begin{align}
    &\Deltil_k \geq \Delta_k \geq 0, \label{eqn:causal-comm}
    \\
    &\Delta_k \leq \Delta(t_k, p_k) \triangleq \frac{ p_k }{ R(t_k) }
    = \frac{b_k}{R_a(t_k)} , \label{eqn:Delta-fun}
  \end{align}
\end{subequations}
where the first condition is that of causal communication and the
second is an upper bound on the communication time. Note that the
actual instantaneous communication rate at $t_k$ is $b_k / \Delta_k$
and we can rewrite~\eqref{eqn:Delta-fun} as
\begin{align*}
  \frac{b_k}{\Delta_k} = \frac{np_k}{\Delta_k} \geq \frac{ np_k }{
    \Delta(t_k,p_k)} = R_a(t) ,
\end{align*}
to realize that $R_a(t)$ is a lower bound on the number of bits
communicated per unit time of all the bits transmitted at time~$t$.
Thus, for example, if $R_a(t) = \infty$, then the packet sent at $t$
is received instantaneously. The packet size $b_k = np_k$ that can be
successfully transmitted starting at $t_k$ is upper bounded as
\begin{subequations}\label{eqn:feas-seq}
  \begin{align}
    p_k \leq \pbar(t_k) , \quad p_k \in \integersnonnegative
  \end{align}
  for all $k \in \integersnonnegative$. We refer to an interval of
  time during which $\pbar = \bbar = 0$ as a \emph{(channel)
    blackout}. In this paper, we assume that the encoder knows the
  functions $t \mapsto R(t)$ and $t \mapsto \pbar(t)$ a priori or
  sufficiently in advance, which we make clear in the sequel.
  
  Since the channel has bounded data capacity and in order to maintain
  synchronization between the encoder and the decoder, we require that
  the encoder does not transmit a packet before a previous packet is
  received by the decoder and the controller updated, i.e.,
  \begin{align}
    t_{k+1} \geq \rtil_k,
  \end{align}
  for all $k \in \integersnonnegative$.
\end{subequations}
We say the \emph{channel is busy} at time $t$ if $t \in [t_k, r_k)$,
for some $k \in \integerspositive$. Finally, we refer to the sequences
of transmission times $\{t_k\} \subset \realnonnegative$, packet sizes
$\{b_k\} \subset \integersnonnegative$, and update times $\{\rtil_k\}
\subset \realnonnegative$ as \emph{feasible} if~\eqref{eqn:Del-cond}
and~\eqref{eqn:feas-seq} are satisfied for every $k \in
\integerspositive$.

\subsection{Encoding and decoding}

We use dynamic quantization for finite-bit transmissions from the
encoder to the decoder. In dynamic quantization, there are two
distinct phases: the zoom-out stage, e.g.,~\citep{DL:03}, during which
no control is applied while the quantization domain is expanded until
it captures the system state at time $r_0 = t_0 \in \realnonnegative$;
and the zoom-in stage, during which the encoded feedback is used to
asymptotically stabilize the system.  We focus exclusively on the
latter, i.e., for $t \geq t_0$. We assume both the encoder and the
decoder have perfect knowledge of the plant system matrices, have
synchronized clocks, and synchronously update their states at update
times $\{\rtil_k\}_{k \in \integerspositive}$. For simplicity, we
assume that at transmission $t_k$ the sensor (encoder) encodes each
dimension of the plant state using $p_k$ bits so that the total number
of bits transmitted is $b_k = np_k$.

The state of the encoder/decoder is composed of the controller state
$\hat{x} \in \real^n$ and an upper bound $d_e \in \realnonnegative$ on
$\Infnorm{x_e}$, where $x_e \triangleq x - \hat{x}$ is the encoding
error. Thus, the actual input to the plant is given by $u(t) = K
\hat{x}(t)$. During inter-update times, the state of the dynamic
controller evolves as
\begin{subequations}\label{eqn:x_hat}
  \begin{align}\label{eqn:xhat_evolve}
    \dot{\hat{x}}(t) = A \hat{x}(t) + B u(t) = \bar{A} \hat{x}(t),
    \quad t \in [\rtil_k, \rtil_{k+1}) .
  \end{align}
  Let the encoding and decoding functions at the $k^{\text{th}}$
  iteration be represented by $q_{E, k}: \mathbb{R}^n \times
  \mathbb{R}^n \mapsto G_k$ and $q_{D, k}: G_k \times \mathbb{R}^n
  \mapsto \mathbb{R}^n$, respectively, where $G_k$ is a finite set of
  $2^{b_k}$ symbols. At $t_k$, the encoder encodes the plant state as
  $z_{E,k} \triangleq q_{E,k}(x(t_k), \hat{x}(t_k^-))$, where
  $\hat{x}(t_k^-)$ is the controller state just prior to the encoding
  time $t_k$, and sends it to the controller. The decoder can decode
  this signal as $z_{D,k} \triangleq q_{D,k}(z_{E,k}, \hat{x}(t_k^-))$
  at any time during $[r_k,\rtil_k]$. At the update time $\rtil_k$,
  the sensor and the controller also update $\hat{x}$ using the jump
  map,
  \begin{align}\label{eqn:x_hat_jmp_nominal}
    \hat{x}(\rtil_k) & = e^{\bar{A} \Deltil_k} \hat{x}(t_k^-) + e^{A
      \Deltil_k} (z_{D,k} - \hat{x}(t_k^-)) \notag
    \\
    & \triangleq q_k(x(t_k), \hat{x}(t_k^-)) ,
  \end{align}
\end{subequations}
where $q_k: \mathbb{R}^n \times \mathbb{R}^n \mapsto \mathbb{R}^n$
represents the quantization that occurs as a result of the finite-bit
coding. We allow the quantization domain, the number of bits and the
resulting quantizer, $q_k$, for each transmission $k \in
\integerspositive$ to be variable.
The evolution of the plant state $x$ and the encoding error~$x_e$ on
the time interval $[\rtil_k, \rtil_{k+1})$ can be written as
\begin{subequations}\label{eq:dynamics-x-xe}
  \begin{align}
    \dot{x}(t) &= \bar{A} x(t) - BK x_e(t),
    \label{eqn:plant_dyn_enc_err}
    \\
    \dot{x}_e(t) &= A x_e(t). \label{eqn:enc_err_dyn}
  \end{align}
\end{subequations}
While the encoder knows the encoding error~$x_e$ precisely, the
decoder can only compute a bound $d_e(t)$ on $\Infnorm{x_e(t)}$ as
follows
\begin{subequations}\label{eqn:coding_scheme}
  \begin{align}
    d_e(t) &\triangleq \Infnorm{e^{A(t-t_k)}} \delta_k ,
    \ t \in [\rtil_k, \rtil_{k+1}), \ k \in \integersnonnegative
    \label{eqn:de_evolve_rk}
    \\
    \delta_{k+1} &= \frac{1}{2^{p_{k+1}}} d_e(t_{k+1}). 
    \label{eqn:de_jump_rk}
  \end{align}
\end{subequations}
One can design a pair of algorithms for the encoder and the decoder to
implement~\eqref{eqn:x_hat_jmp_nominal} in a manner that they maintain
consistent $\hat{x}(t)$ and $d_e(t)$ signals for~$t \geq t_0$
(see~\citep{PT-JC:16-tac} for example). For the sake of brevity, we do
not present these algorithms here and it suffices to say that
$\Infnorm{x_e(t)} \leq d_e(t)$ for all $t \geq t_0$ if
$\Infnorm{x_e(t_0)} \leq d_e(t_0)$.

\subsection{Control objective}\label{sec:con-obj}

We measure the performance of the closed-loop system through a
Lyapunov function as follows.  Given an arbitrary symmetric positive
definite matrix $Q \in \real^{n \times n}$, let $P$ be the unique
symmetric positive definite matrix that satisfies the Lyapunov
equation
\begin{align}\label{eq:Lyap-eq}
  P \bar{A} + \bar{A}^T P = - Q.
\end{align}
Define $x \mapsto V(x) = x^T P x$ and let
\begin{equation}\label{eqn:Vd_t}
  V_d(t) = V_d(t_0) e^{ -\beta (t - t_0)} ,
\end{equation}
with $\beta > 0$, be the desired \emph{control performance}. We assume
that
\begin{align}\label{eq:W}
  W & \triangleq \frac{\lambda_m(Q)}{\lambda_M(P)} - a \beta > 0 ,
\end{align}
with $a > 1$ an arbitrary constant. Assumption~\eqref{eq:W} is
sufficient to guarantee a convergence rate faster than $\beta$ for the
dynamics~\eqref{eqn:plant_dyn} under the continuous-time and
unquantized feedback $u(t) = K x(t)$.

Given the system and the communication channel model above, our
objective is to design an event-triggered communication and control
strategy that ensures the exponential stability of the origin.
Formally, we seek to synthesize an event-triggered control strategy
that recursively determines the sequences of transmission times
$\{t_k\}_{k \in \integerspositive}$ and update times $\{\rtil_k\}_{k
  \in \integerspositive}$, along with a coding scheme for messages and
a rule to determine the number of bits $\{b_k\}_{k \in
  \integerspositive}$ to be transmitted, so that
\begin{align*}
  V(x(t)) \leq V_d(t) ,
\end{align*}
holds for all $t \geq t_0$. This objective is especially challenging
given the time-varying nature of the communication channel and the
possibility of intermittent blackouts.

\section{Performance- and channel-trigger functions}


In order to achieve the control objective of Section~\ref{sec:con-obj}
with opportunistic transmissions, we need a performance-trigger
function that tells us how close the system state is to violating the
convergence requirement. Bounded precision quantization further
requires us to keep track (through a channel-trigger function) of the
number of bits required at any moment to guarantee performance at
least for a certain period of time. Threshold crossings of these two
functions form the primary basis of our event-triggering
mechanism. Further, in order to take care of communication delays, the
triggering mechanism instead uses guaranteed upper bounds on the
performance and channel-trigger functions up to the maximum possible
communication delay for the current channel state. In this section, we
describe each of these components, thus laying the groundwork to deal
with time-varying communication channels and blackouts.

\subsection{Performance-trigger function}

We define the \emph{performance-trigger} function as the ratio of the
quadratic Lyapunov function $V$ and the desired performance~$V_d$,
\begin{equation}\label{eqn:bt_def}
  \triggerPf(t) \triangleq \frac{V(x(t))}{V_d(t)} .
\end{equation}
Note that the control objective is to maintain $\triggerPf(t) \le 1$
at all times. This is why, in general, it is of interest to
characterize the open-loop evolution of the performance-trigger
function.  The next result provides an upper bound on the value of
$\triggerPf$ in the future as a function of the information available
now.

\begin{lemma}\longthmtitle{Upper bound on open-loop evolution of
    performance-trigger function~\citep{PT-JC:16-tac}} \label{lem:bound-b}
  Given $t_k \in \realpositive$ such that $\triggerPf(t_k) \leq 1$,
  then
  \begin{align*}
    \triggerPf(\tau + t_k) \leq \triggerPfbar(\tau, \triggerPf(t_k),
    \epsilon(t_k)),
  \end{align*}
  for $\tau \ge 0$, where
  \begin{align}
    &\epsilon(t) \triangleq \frac{d_e(t)}{c \sqrt{V_d(t)}}, \quad
    \triggerPfbar(\tau, h_0, \epsilon_0) \triangleq \frac{ f_1(\tau, h_0,
      \epsilon_0) }{ f_2(\tau) } , \label{eqn:eps-btild-def}
    \\
    &f_1(\tau, h_0, \epsilon_0) \triangleq h_0 + \frac{W \epsilon_0 }{
      w + \mu } ( e^{(w + \mu) \tau} - 1 ), \quad f_2(\tau)
    \triangleq e^{w \tau} , \notag
    \\
    &c \triangleq \frac{W \sqrt{ \lambda_m(P)} }{ 2 \sqrt{n}
      \Enorm{PBK} } , \; w \triangleq
    \frac{\lambda_m(Q)}{\lambda_M(P)} - \beta >0, \; \mu
    \triangleq \Enorm{A} + \frac{\beta}{2} . \notag
  \end{align}
\end{lemma}

This result motivates the definition of the function
\begin{align*}
  \Gamma_1(h_0, \epsilon_0) \triangleq \min \{ \tau \geq 0 :
  \triggerPfbar(\tau, h_0, \epsilon_0) = 1, \ \frac{\mathrm{d}
    \triggerPfbar}{\mathrm{d} \tau} \geq 0 \} ,
\end{align*}
as a lower bound on the time it takes $\triggerPf$ to evolve to $1$
starting from $\triggerPf(t_k) = h_0$ with $\epsilon(t_k) =
\epsilon_0$. The following result captures some useful properties of
this function.

\begin{lemma}\longthmtitle{Properties of the function
    $\Gamma_1$~\citep{PT-JC:16-tac}}\label{lem:Gamma1_prop}
  The following holds true,
  \begin{enumerate}
  \item $\Gamma_1(1, 1) > 0$.
  \item If $h_1 \geq h_0$ and $\epsilon_1 \geq \epsilon_0$, then
    $\Gamma_1(h_0, \epsilon_0) \geq \Gamma_1(h_1, \epsilon_1)$. In
    particular, if $h_0 \in [0, 1]$, then $\Gamma_1(h_0, \epsilon_0)
    \geq \Gamma_1(1, \epsilon_0)$.
  \item For $T>0$, if $h_0 \in [0, 1]$ and
    \begin{equation}\label{eqn:rho_def}
      \epsilon_0 \leq \rhofun{T}{h_0} \triangleq \frac{ (w + \mu) 
        ( 1 - h_0 )  }{ W ( e^{ (w+\mu) T } - 1 ) } + 1 ,
    \end{equation}
    then $\Gamma_1(h_0, \epsilon_0) \geq \min \{ \Gamma_1(1, 1), T
    \}$.
  \item For $T > 0$ and $h_0 \in [0, 1]$,
    \begin{align*}
      \Gamma_1(h_0, \epsilon_0) \geq T \Longleftrightarrow \triggerPfbar(T,
      h_0, \epsilon_0) \leq 1.
    \end{align*}
    The statement with strict inequalities is also true.
  \end{enumerate}
\end{lemma}



\subsection{Channel-trigger function}

We define the \emph{channel-trigger} function
\begin{equation}\label{eqn:h2_def}
  \triggerCh(t) \triangleq  \frac{ \epsilon(t) }{
    \rhofun{T}{\triggerPf(t)} } ,
\end{equation}
where $T > 0$ is a fixed design parameter. The channel-trigger
function $\triggerCh$ depends on the bound on the encoding error $d_e$
through $\epsilon$. Note that the channel-trigger function
$\triggerCh$ through its dependence on $d_e$, which evolves
as~\eqref{eqn:coding_scheme}, also jumps at the update times
$\rtil_k$. Lemma~\ref{lem:Gamma1_prop}(iii) implies that for any time
$s_0 \geq t_0$, if $\triggerCh(s_0) \leq 1$, then $\triggerPf(t) \leq
1$ for at least $t \in [s_0, s_0 + \min \{ T, \Gamma_1(1,1) \} )$ even
without any transmissions or receptions. Thus, assuming that the
communication delays are smaller than $\min \{ T, \Gamma_1(1,1) \}$, a
transmission strategy ( $ \{t_k\}_{k \in \integerspositive}$ and
$\{b_k\}_{k \in \integerspositive}$ such that $b_k = n p_k$) is to
ensure that, for each $k$, $\triggerCh(\rtil_k) \leq 1$ so that
$\Gamma_1( \triggerPf( \rtil_k ), \epsilon( \rtil_k ) ) \geq \min \{
T, \Gamma_1(1,1) \}$. Thus, we now require an upper bound on the
open-loop evolution of $\triggerCh$, which is provided in the
following result.  Its proof follows from the definitions of
$\epsilon$ and $\rho_{T}$ in~\eqref{eqn:eps-btild-def}
and~\eqref{eqn:rho_def}, respectively, and the evolution of $d_e$
described in~\eqref{eqn:coding_scheme}.

\begin{lemma}\longthmtitle{Upper bound on the channel-trigger function
    at the update times $\rtil_k$}\label{lem:bound-triggerCh}
  If $t_k \in \realpositive$ is such that $\triggerPf(t_k) \in [0,
  1]$, then
  \begin{align}\label{eqn:hch-evolve}
    \triggerCh(\rtil_k) \leq \triggerChbar( \rtil_k - t_k,
    \triggerPf(t_k), \epsilon(t_k), p_k ) ,
    \end{align}  
    where $b_k = n p_k$ bits are transmitted at $t_k$ and
  \begin{align}\label{eqn:hbar_def}
    \triggerChbar(\tau, h_0, \epsilon_0, p) \triangleq
    \frac{\Infnorm{e^{A \tau}} e^{\frac{\beta}{2} \tau} \epsilon_0 }{
      \rhofun{T}{\triggerPfbar(\tau, h_0, \epsilon_0)} } \cdot \frac{
      1 }{ 2^p }.
  \end{align}
\end{lemma}

Note that for $t, t+\tau \in [\rtil_k, t_{k+1})$, for any $k \in
\integersnonnegative$, we have $\triggerCh(t+\tau) \leq
\triggerChbar(\tau, \triggerPf(t), \epsilon(t), 0)$.

Now, analogous to~$\Gamma_1$, we define
\begin{align}\label{eqn:Gamma2_tilde_def}
  \Gamma_2(b_0, \epsilon_0, p) &\triangleq \min \{ \tau \geq 0 :
  \triggerChbar(\tau, b_0, \epsilon_0, p) = 1 \} ,
\end{align}
which essentially is an upper bound on the communication delay
$\rtil_k - t_k$, for which we can still guarantee $\triggerCh(\rtil_k)
\leq 1$. Given the interpretation of $\Gamma_2$, one of the conditions
in our event-triggering rule would be to check if $\Gamma_2$ is less
than a maximum communication delay. The next result provides a way to
check this in real time.

\begin{lemma}\longthmtitle{Algebraic condition to check value
    of~$\Gamma_2$~\citep{PT-JC:16-tac}} \label{lem:Gamma2-T_sign}
  Let $T^\circ > 0$. For any $h_0 \in [0, 1]$ and $\epsilon_0 \in [0,
  \rhofun{T}{h_0}]$, $\Gamma_2(h_0, \epsilon_0, p) > T^\circ$ if and
  only if $\triggerChbar(T^\circ, h_0, \epsilon_0, p) < 1$. Further,
  the statement with equalities is also true.
\end{lemma}

The following result provides a lower bound for $\Gamma_2$
uniform in its first two arguments. This bound will be useful in our
event-triggered design later.

\begin{lemma}\longthmtitle{Lower bound on
    $\Gamma_2$}\label{lem:Tstar-unif-lbnd-Gamma2}
  If $\epsilon_0 \in [0, \rhofun{T}{h_0}]$ then $\Gamma_2(h_0,
  \epsilon_0, p) \geq T^*(p)$ with
  \begin{align*}
    T^*(p) &\triangleq \min \{ \tau \geq 0: g( \tau, p ) = 1 \} ,
    \\
    g(\tau, p) &\triangleq \frac{\Infnorm{e^{A \tau}}
      e^{\frac{\beta}{2} \tau} }{ 2^p } \cdot \frac{ e^{ (w+\mu) T
      } - 1 }{ e^{ (w+\mu) T } - e^{(w + \mu) \tau} } .
  \end{align*}
\end{lemma}
\begin{pf}
  From~\eqref{eqn:bt_def} and~\eqref{eqn:rho_def}, we have
  \begin{align*}
    &\rhofun{T}{\triggerPfbar(\tau, h_0, \epsilon_0)}
    \\
    &= \frac{ (w + \mu) ( 1 - e^{-w \tau} ( h_0 + \frac{W
        \epsilon_0 }{ w + \mu } ( e^{(w + \mu) \tau} - 1 ) ) )
    }{ W ( e^{ (w+\mu) T } - 1 ) } + 1
    \\
    &= \rhofun{T}{e^{-w \tau} h_0} - \frac{ e^{(w + \mu) \tau} - 1
    }{ e^{ (w+\mu) T } - 1 } e^{-w \tau} \epsilon_0
    \\
    &\geq \rhofun{T}{e^{-w \tau} h_0} \frac{ e^{ (w+\mu) T } -
      e^{(w + \mu) \tau} }{ e^{ (w+\mu) T } - 1 } ,
  \end{align*}
  where the inequality follows from the assumption that $\epsilon_0
  \leq h_0$. Now, substituting this lower bound
  in~\eqref{eqn:hbar_def} and noting the fact that $\rhofun{T}{e^{-w
      \tau} h_0} \geq \rhofun{T}{h_0}$ gives
  \begin{align*}
    \triggerChbar(\tau, h_0, \epsilon_0, p) \leq g(\tau, p).
  \end{align*}
  The claim now follows from the
  definition~\eqref{eqn:Gamma2_tilde_def}. \qed
\end{pf}

\myclearpage
\section{Characterization of the data capacity}\label{sec:channel}

Our study of data capacity here is motivated by the need of the
encoder to know how much data can be transmitted successfully before a
channel blackout.

\subsection{Data capacity}

We denote the number of bits (data) \emph{communicated} (the data
transmitted by the encoder and completely received by the decoder)
during the time interval $[\tau_1, \tau_2]$ under the feasible
sequences $\{t_k\}$, $\{p_k\}$, and $\{\Deltil_k\}$ (that
satisfy~\eqref{eqn:Del-cond} and~\eqref{eqn:feas-seq}) as
\begin{equation*}
  \DataWithSeq( \tau_1, \tau_2, \{t_k\}, \{\Deltil_k\}, \{p_k\} ) \triangleq
  n \sum_{k= \uline{k}_{\tau_1} }^{ \overline{k}_{\tau_2} } p_k,
\end{equation*}
where $ \uline{k}_{\tau_1} = \min \{k: t_k \geq \tau_1 \}$ and $
\overline{ k}_{\tau_2} = \max \{k: t_k + \Deltil_k \leq \tau_2 \}$.
Notice that we count only the bits that are transmitted and also
received (\emph{communicated}) during $[\tau_1, \tau_2]$. We define
the data capacity during the time interval $[\tau_1, \tau_2]$ as the
maximum data that can be communicated during the time interval under
\emph{all} possible communication delays, i.e.,
\begin{equation*}
  \Datacap{\tau_1}{\tau_2} \triangleq \max_{ \substack{\{t_k\},
      \{p_k\} \\ \text{s.t. \eqref{eqn:feas-seq} holds} \\ \forall
      \Delta_k \leq \Delta(t_k, p_k)} } 
  \DataWithSeq( \tau_1, \tau_2, \{t_k\}, \{\Delta_k\}, \{p_k\} ).
\end{equation*}
Notice that to maximize the data communicated, it must be that
$\rtil_k = r_k$ ($\Deltil_k = \Delta_k$) for all $k \in
\integerspositive$. This explains the fact that only the sequences
$\{t_k\}$ and $\{p_k\}$ are the optimization variables. Next, notice
that maximization under \emph{all} possible communication delays
($\Delta_k \leq \Delta(t_k, p_k)$) is the same as maximization under
maximum communication delays ($\Delta_k = \Delta(t_k, p_k)$). Thus,
the definition of the data capacity reduces to
\begin{equation}\label{eqn:Datacap-def}
  \Datacap{\tau_1}{\tau_2} \triangleq \max_{ \substack{\{t_k\},
      \{p_k\} \\ \text{s.t. \eqref{eqn:feas-seq} holds} } } 
  \DataWithSeq( \tau_1, \tau_2, \{t_k\}, \{\Delta(t_k, p_k)\}, \{p_k\} ).
\end{equation}

Note that a greedy approach does not necessarily maximize the
communicated data. 
In general, the precise computation of $\Datacap{\tau_1}{\tau_2}$
involves solving an integer program with non-convex feasibility
constraints. Given the difficulty of solving this problem, we seek a
class of channel functions $R$ and $\bbar$ that are meaningful and yet
simple enough to efficiently compute a lower bound for the data
capacity. To this end, we make the following observation.

\begin{lemma}\longthmtitle{Data capacity under constant
    communication rate}\label{lem:Dcap-constR}
  Suppose $\forall t \in [\tau_1, \tau_2]$ (i) $R(t) = R \geq 0$ and
  (ii) $\pbar(t) \geq 1$ (no blackouts). Then,
  $\Datacap{\tau_1}{\tau_2} = n \floor{R(\tau_2 - \tau_1)}$.
\end{lemma}

The proof of Lemma~\ref{lem:Dcap-constR} follows directly by noting
that an optimal solution can be constructed by choosing $p_k = 1$ and
$t_{k+1} = \rtil_k = r_k$ for all $k \in \integersnonnegative$.
Motivated by this result, we assume in the sequel that the channel
function $R$ is piecewise constant so that the problem of finding a
reasonable lower bound on $\Datacap{\tau_1}{\tau_2}$ is tractable
while also ensuring that the overall problem is meaningful. Note that
any given $R$ can be approximated to an arbitrary degree of accuracy
by a piecewise constant function. In addition, according
to~\eqref{eqn:Delta-fun}, $R$ is a lower bound on the instantaneous
communication rate and it is quite reasonable to assume it is
piecewise constant. Also, note that $\pbar$ takes integer values and
hence by its nature is always piecewise constant. Specifically, we
assume that
\begin{subequations}\label{eqn:pwc-R-pbar}
  \begin{align}
    R(t) &= R_j, \quad \forall t \in ( \theta_j, \theta_{j+1} ]
    \\
    \pbar(t) &= \pibar_j, \quad \forall t \in ( \theta_j, \theta_{j+1}
    ] \label{eqn:pwc-pbar}
  \end{align}
\end{subequations}
where $\{\theta_j\}_{j=0}^{\infty}$ is a strictly increasing sequence
of time instants and $\pibar_j \in \integersnonnegative$ for each
$j$. We also denote $T_j \triangleq \theta_{j+1} - \theta_j$ as the
length of the $j^{\text{th}}$ time slot $I_j \triangleq ( \theta_j,
\theta_{j+1} ]$. Again note that identical $\{\theta_j\}$ sequences
for $R$ and $\pbar$ is not a restriction because one can always refine
the sequence $\{\theta_j\}$. In order to concisely express the
constraints in the optimization problem~\eqref{eqn:Datacap-def} we
assume, without loss of generality, that $\tau_1 = \theta_{j_0}$ and
$\tau_2 = \theta_{j_f}$, for some $j_0, j_f \in
\integersnonnegative$. Finally, we choose left-open intervals in our
model~\eqref{eqn:pwc-R-pbar} since it provides a slight technical
advantage in lowering the gap between the optimal and our sub-optimal
solutions.

\subsection{Formulation as an allocation problem}

Here we show that, for piecewise constant channel functions, we can
think of the computation of $\Datacap{\theta_{j_0}}{\theta_{j_f}}$ as
an allocation problem: that of allocating the number of bits
$\{n\phi_j\}$, with $\phi_j \in \integersnonnegative$, to be
transmitted in the time slots $\{I_j\}$ for $j \in \Nzf \triangleq \{
j_0, \ldots, j_{f}-1\}$. For convenience, we let $\phizf \triangleq
(\phi_{j_0}, \ldots, \phi_{j_f - 1})$.
Given $\phizf$, the sequences $\{t_k\}$ and $\{p_k\}$ are determined
so that transmissions start at the earliest possible time in $I_j$ and
the channel is not idle until all the allocated bits $\phi_j$ are
received, i.e., $t_{k+1} = \rtil_k = r_k = \Delta(t_k, p_k)$ during
$I_j$ and $\{p_k\}$ during $I_j$ is any sequence that respects the
channel upper bound $\pibar_j$ and adds up to $\phi_j$.  Given this
correspondance, our forthcoming discussion 
focuses on expressing the constraints in the optimization problem in
terms of the $\phi$ variables. In the sequel, a standing constraint is
that $\phi_j \in \integersnonnegative$ for each $j$, unless we mention
otherwise.

\emph{Maximum bits that may be transmitted:} First, we present the
constraint that describes the maximum bound on the number of bits that
may be transmitted in each slot $I_j$. Note that according to
Lemma~\ref{lem:Dcap-constR}, in the time slot $I_j$, $n \floor{R_j
  T_j}$ bits could be transmitted and received within $\floor{R_j T_j}
/ R_j \leq T_j$ units of time. In addition, $n \pibar_j$ more bits
could be transmitted during the closed interval $[ \ \floor{R_j T_j},
\theta_{j+1} ]$, though these bits are received only in subsequent
time slots. Thus, we have for each $j \in \Nzf$
\begin{align}\label{eqn:max-bit-bound-Ij}
  n\phi_j \leq
  \begin{cases}
    n R_j T_j + n\pibar_j, \quad &\text{if } \pibar_j > 0 
    \\
    0, \quad &\text{if } \pibar_j = 0 
  \end{cases}
\end{align}
where in the first case we have used the fact that $\phi_j \in
\integersnonnegative$ to avoid the use of the floor function.

\emph{Reduced channel availability in a time slot due to prior
  transmissions:} As noted above, if $\phi_j > \floor{R_j T_j}$, then
these bits take up some of the time in $I_{j+1}$ and possibly even
subsequent slots. Thus, effectively the time available in $I_{j+1}$
and consequently the upper bound on $\phi_{j+1}$ is reduced. Moreover,
in general, the number of bits transmitted in $I_j$ has an effect on
the number that could be transmitted in all subsequent intervals
either directly or indirectly. Thus, for each $j_1, j \in \Nzf$, we
introduce
%
\begin{align}\label{eqn:T-j1-j}
  \Tbar_{j_1, j}(\phizf) &\triangleq \Big( T_j - \sum_{ i=j_1}^{j-1}
  \big( \frac{ \phi_i }{ R_i } - T_i \big) \Big) \notag
  \\
  &= \theta_{j+1} - \theta_{j_1} - \sum_{ i=j_1}^{j-1} \frac{ \phi_i
  }{ R_i } .
\end{align}
As we shall see in the following lemma, these functions determine the
available time in slot $I_j$ given $\phizf$.

\begin{lemma}\longthmtitle{Available time in slot $I_j$}\label{lem:Tbar}
  Let $\Tbar_j(\phizf)$ be the time available in the slot $I_j$ given
  the allocation $\phizf$. Then,
  \begin{equation*}
    \Tbar_j(\phizf) = \Bsat{ \min_{j_1 \in \Nzf} \{ \Tbar_{j_1, j}(\phizf),
      T_j \} } .
  \end{equation*}
\end{lemma}
\begin{pf}
  Observe that for any $j_1, j \in \Nzf$, $\theta_{j+1} -
  \theta_{j_1}$ is the total time in the slots $j_1$ to $j$, while
  $\sum_{ i=j_1}^{j-1} \frac{ \phi_i }{ R_i }$ is the total time taken
  by the bits transmitted in slots $j_1$ to $j-1$. Thus,
  $\losat{\Tbar_{j_1, j}(\phizf)}$ is an upper bound on the time
  available for transmission in the slot $I_j$. Now, let
  \begin{equation*}
    j_2 = \max \{ i \in \integersnonnegative \intersect [j_0, j-1] :
    \Tbar_i(\phizf) = T_i \}
  \end{equation*}
  Then clearly, $\{\phi_i\}_{i = j_2}^{j-1}$ is sufficient to
  determine $\Tbar_j(\phizf)$. Next, for the allocation $\phizf$, the
  bits transmitted during the time slots $I_i$ for $i \in \{ j_2, j-1
  \}$ are received by $\theta_{j_2} + \sum_{ j=j_2}^{j-1} \tfrac{
    \phi_j }{ R_j }$ and thus in deed $\Tbar_j(\phizf) = \losat{ \min
    \{ \Tbar_{j_2, j}(\phizf), T_j \} }$. Finally, for each $j_1 \in
  \integersnonnegative \intersect [j_0, j_2-1]$, $\Tbar_{j_1,
    j}(\phizf) \geq \Tbar_{j_2, j}(\phizf)$, which proves the
  result. \qed
\end{pf}

As a consequence of Lemma~\ref{lem:Tbar}, for each $j \in \Nzf$ and
$j_1 \in \integersnonnegative \intersect [j_0, j-1]$, consider the
constraints
\begin{subequations}\label{eqn:case-constraints}
  \begin{equation}\label{eqn:coupled-max-bit-bound}
    n \phi_j \leq \begin{cases}
      n R_j \Tbar_{j_1, j}(\phizf) + n \pibar_j,  &\text{if }
      \Tbar_{j_1, j}(\phizf) > 0
      \\
      0 &\text{otherwise}
    \end{cases}
  \end{equation}
  which we obtain using the same reasoning as
  in~\eqref{eqn:max-bit-bound-Ij} with $T_j$ replaced by $\Tbar_{j_1,
    j}(\phizf)$. Note that if $\Tbar_{j_1, j}(\phizf) \geq T_j$, then
  the constraint~\eqref{eqn:coupled-max-bit-bound} is weaker
  than~\eqref{eqn:max-bit-bound-Ij} and hence inactive. For
  $\Tbar_{j_1, j}(\phizf) \in (0,T_j)$, the constraint reflects the
  reduced available time in the time slot $I_j$ and if $\Tbar_{j_1,
    j}(\phizf) \leq 0$, for some $j_1 \in \integersnonnegative
  \intersect [j_0, j-1]$, then it corresponds to the case when the
  channel is busy for the whole of the time slot $I_j$
  ($\Tbar_j(\phizf) = 0$). Thus~\eqref{eqn:coupled-max-bit-bound}
  accurately reflects the effect of possibly reduced available time
  during the slot $I_j$ due to prior transmissions.

  \emph{Counting only the bits transmitted and received during $[\theta_{j_0},
    \theta_{j_f}]$:} Finally, since in the computation of
  $\Datacap{\theta_{j_0}}{\theta_{j_f}}$, we are interested in the
  maximum number of bits that can be communicated (transmitted and
  received) during the time interval, we also require that any bits
  transmitted during the slot $I_j$ are received before
  $\theta_{j_f}$, i.e.,
  \begin{equation*}
    \frac{ \phi_j }{ R_j } \leq \begin{cases}
      \Tbar_j(\phizf) + \theta_{j_f} - \theta_{j+1}, &\text{if }
      \Tbar_j(\phizf) > 0
      \\
      0, &\text{otherwise}.
    \end{cases}
  \end{equation*}
  Using the definition of $\Tbar_j(\phizf)$, this can be rewritten
  giving the following constraints for each $j \in \Nzf$ and $j_1 \in
  \integersnonnegative \intersect [j_0, j]$
  \begin{equation}\label{eqn:receive-before-thetaf}
    \frac{ \phi_j }{ R_j } \leq \begin{cases}
      \Tbar_{j_1, j}(\phizf) + \theta_{j_f} - \theta_{j+1}, &\text{if
      } \Tbar_{j_1, j}(\phizf) > 0
      \\
      0, &\text{otherwise} .
    \end{cases}
  \end{equation}
\end{subequations}
Then, the data capacity is given as
\begin{equation}\label{eqn:ch-cap-reform}
  \Datacap{\theta_{j_0}}{\theta_{j_f}} = \max_{ \substack{ \phi_j \in
      \integersnonnegative, \forall j \in \Nzf
      \\
      \text{s.t. \eqref{eqn:max-bit-bound-Ij}, 
        \eqref{eqn:case-constraints} hold} } } 
  \; n \sum_{j=j_0}^{j_f - 1} \phi_j.
\end{equation}
Ignoring the fact that this is an integer program, the
constraints~\eqref{eqn:case-constraints} still make the problem
combinatorial.

\subsection{Efficient approximation of data
  capacity}\label{sec:efficient-approx}

The following result is the basis for the construction of a
sub-optimal and efficient solution to the
problem~\eqref{eqn:ch-cap-reform}.

\begin{lemma}\longthmtitle{Bound on ``channel
    variation''}\label{lem:R-pi-T}
  If there exists $J \in \integersnonnegative$ such that
  \begin{equation}\label{eqn:R-pi-T}
    \displaystyle \frac{ \pibar_j }{ R_j } < \sum_{i=j+1}^{i=j+1+J}
    T_i, \quad \forall j \in \Nzf ,
  \end{equation}
  then, for any $j \in \Nzf$, any bits transmitted in time slot $I_j$
  would be received strictly before the end of the slot $I_{j+1+J}$.
\end{lemma}
\begin{pf}
  The term $\pibar_j/R_j$ is the time it takes a packet of size up to
  $n \pibar_j$ bits transmitted during $I_j$ to reach the
  decoder. Thus, the claim follows by noting that any bits transmitted
  during $I_j$ would be received before $t =~ \theta_{j+1} +
  (\pibar_j/R_j)$. \qed
\end{pf}

Lemma~\ref{lem:R-pi-T} relates the three sequences of parameters,
$\{R_j\}$, $\{\pibar_j\}$ and $\{T_j\}$, that define the channel state
at any given time. The result may be interpreted as the imposition of
a bound on how often there is a change in the channel state as
measured by the time slot lengths $T_j$.  The parameter~$J$ may be
interpreted as a uniform upper bound on the number of consecutive time
slots that may be fully occupied due to a prior transmission.

\subsubsection{Guaranteed channel availability in each time slot}

The case of $J = 0$ is of special interest and will be addressed
next. This case is interesting because the
constraints~\eqref{eqn:case-constraints} reduce to a simpler form, as
presented in the following result, and using which we can compute a
good sub-optimal solution subsequently.

\begin{lemma}\longthmtitle{Data capacity in the case of $J =
    0$}\label{lem:optim-J0}
  Suppose the channel is such that $J = 0$ for all $j \in \Nzf$. Then,
  the constraints~\eqref{eqn:coupled-max-bit-bound} reduce to
  \begin{subequations}\label{eqn:case-constraints-J0}
    \begin{align}\label{eqn:J0-coupled-max-bits}
      n \phi_j + n R_j \sum_{ i=j_1}^{j-1} \frac{ \phi_i }{ R_i } \leq
      n R_j ( \theta_{j+1} - \theta_{j_1} ) + n \pibar_j ,
    \end{align}
    for each $j \in \Nzf$ and $j_1 \in \integersnonnegative \intersect
    [j_0, j-1]$ while the
    constraints~\eqref{eqn:receive-before-thetaf} reduce to
    \begin{equation}\label{eqn:J0-eqn:receive-before-thetaf}
      \sum_{ i=j_1}^{j_f-1} \frac{ \phi_i }{ R_i } \leq \theta_{j_f} -
      \theta_{j_1} ,
    \end{equation}
    for each $j_1 \in \integersnonnegative \intersect [j_0, j_f-1]$.
  \end{subequations}
  The data capacity is
  \begin{equation}\label{eqn:ch-cap-J0}
    \Datacap{\theta_{j_0}}{\theta_{j_f}} = \max_{ \substack{ \phi_j \in
        \integersnonnegative, \forall j \in \Nzf
        \\
        \text{s.t. \eqref{eqn:max-bit-bound-Ij}, 
          \eqref{eqn:case-constraints-J0} hold} } } 
    \; n \sum_{j=j_0}^{j_f - 1} \phi_j.
  \end{equation}
\end{lemma}
\begin{pf}
  Indeed, if $J = 0$ then for each $j$ and $j_1 \in \Nzf$,
  $\Tbar_{j_1, j}(\phizf) > 0$ and hence $\Tbar_j > 0$ also. Thus, the
  constraints~\eqref{eqn:coupled-max-bit-bound} reduce to $ n \phi_j
  \leq n R_j \Tbar_{j_1, j}(\phizf) + n \pibar_j$, which after
  using~\eqref{eqn:T-j1-j} give
  us~\eqref{eqn:J0-coupled-max-bits}. Note that
  Lemma~\ref{lem:R-pi-T}, with $J = 0$, guarantees that the
  constraints~\eqref{eqn:receive-before-thetaf} are satisfied for all
  $j \in \{j_0, \ldots, j_f - 2\}$, while for $j_f - 1$
  ~\eqref{eqn:receive-before-thetaf} reduce to
  \begin{equation*}
    \frac{ \phi_{j_f - 1} }{ R_{j_f - 1} } \leq \Tbar_{j_1, j}(\phizf) ,
  \end{equation*}
  which by expanding and rearranging the terms, we get the
  constraints~\eqref{eqn:J0-eqn:receive-before-thetaf}. Data
  capacity~\eqref{eqn:ch-cap-J0} follows
  from~\eqref{eqn:ch-cap-reform} and the equivalence
  of~\eqref{eqn:case-constraints}
  and~\eqref{eqn:case-constraints-J0}. \qed
\end{pf}

Note that for $J=0$ all the constraints, \eqref{eqn:max-bit-bound-Ij}
and \eqref{eqn:case-constraints-J0} are linear, though $\phi_j$ are
still restricted to be integers. This brings us to the next result.


\begin{proposition}\longthmtitle{A sub-optimal solution and
    quantification of sub-optimality in the case of $J =
    0$} \label{prop:suboptim-J0}
  Suppose the channel is such that $J = 0$ for all $j \in \Jc = \{j_0,
  \ldots, j_f\}$. Let $\Subcap{\theta_{j_0}}{\theta_{j_f}} \triangleq n
    \sum_{j=j_0}^{j_f - 1} \phi^N_j $ where
  \begin{align}
    &\phi^N \triangleq \floor{ \phi^r } \triangleq ( \floor{
      \phi^r_{j_0} }, \ldots, \floor{ \phi^r_{j_f - 1} } )
    , \label{eqn:phi-N}
    \\
    &\phi^r = \!\! \argmax_{ \substack{ \phi_j \in \realnonnegative, \
        \forall j \in \Nzf
        \\
        \text{s.t. \eqref{eqn:max-bit-bound-Ij},
          \eqref{eqn:case-constraints-J0} hold} } } \;
    \sum_{j=j_0}^{j_f - 1} \phi_j \notag .
\end{align}
Then $\phi^N$ is a sub-optimal solution to~\eqref{eqn:ch-cap-J0}, i.e.
$\Subcap{\theta_{j_0}}{\theta_{j_f}} \leq
\Datacap{\theta_{j_0}}{\theta_{j_f}}$ and
\begin{align*}
  \Datacap{\theta_{j_0}}{\theta_{j_f}} &-
  \Subcap{\theta_{j_0}}{\theta_{j_f}}
  \\
  &\leq n \card{ \{j \in \integersnonnegative \intersect [j_0, j_{f-1}]
    : \pibar_j > 0 \} } .
\end{align*}
\end{proposition}
\begin{pf}
  Clearly, $\phi^N$ satisfies the
  constraints~\eqref{eqn:max-bit-bound-Ij}
  and~\eqref{eqn:case-constraints-J0} since $\phi^r$ does and for each
  $j$, $\phi^N_j \leq \phi^r_j$ and $\phi^N \in \integersnonnegative$.
  Thus, $\phi^N$ is a sub-optimal solution
  to~\eqref{eqn:ch-cap-J0}. The sub-optimality bound follows from the
  fact that for any $a \in \real$, $( a - \floor{a} ) \in [0,
  1)$. \qed
\end{pf}


\subsubsection{No guaranteed channel availability} 

If $J > 0$, we forgo optimality 
in favor of an easily computable lower bound of the data
capacity. With a slight abuse of notation, we let
\begin{equation*}
  \phi^N_j = \floor{ R_j( \theta_{j+1} - \theta_j ) }, \quad j \in
  \integersnonnegative ,
\end{equation*}
which is the number of bits that can be communicated (transmitted and
received) during the time slot $I_j = [\theta_j,
\theta_{j+1})$. Hence, $\{ \phi_j^N \}_{j \in \integersnonnegative}$ is
a feasible solution and, again with an abuse of notation, we denote
\begin{equation*}
  \Subcap{\theta_{j_0}}{\theta_{j_f}} \triangleq n \sum_{j=j_0}^{j_f
    - 1} \phi^N_j ,
\end{equation*}
which is a sub-optimal lower bound of the data capacity.

\subsection{Computing data capacity in real time}

As mentioned earlier, we want the encoder to compute a lower bound for
the data capacity up to the end of the next blackout
period. However, the computation of $\Subcap{\tau_1}{\tau_2}$ in the
case of $J = 0$ involves solving a linear program and hence may not be
suitable for real-time computation. Thus, given
$\Datacap{\theta_{j_0}}{\theta_{j_f}}$ (or
$\Subcap{\theta_{j_0}}{\theta_{j_f}}$), we propose a simpler procedure
to compute a lower bound on $\Datacap{t}{\theta_{j_f}}$ (or
$\Subcap{t}{\theta_{j_f}}$) for any $t \in [\theta_{j_0},
\theta_{j_0+1})$. We present the procedure in the following result.

\begin{proposition}\longthmtitle{Real-time computation of
    data capacity}\label{prop:Dcap-realtime}
  Let $\phi^*$ (or $\phi^N$) be any optimizing solution to
  $\Datacap{\theta_{j_0}}{\theta_{j_f}}$ (or
  $\Subcap{\theta_{j_0}}{\theta_{j_f}}$). Let
  \begin{align}
    \hatDcap{t}{\theta_{j_f}} &\triangleq \losat{ n \left \lfloor
        \phi^*_{j_0} - R_{j_0} ( t - \theta_{j_0} ) \right \rfloor } +
    n \sum_{j=j_0 + 1}^{j_f - 1} \phi^*_j
    \\
    \hatScap{t}{\theta_{j_f}} &\triangleq \losat{ n \left \lfloor
        \phi^N_{j_0} - R_{j_0} ( t - \theta_{j_0} ) \right \rfloor } +
    n \sum_{j=j_0 + 1}^{j_f - 1} \phi^N_j , \label{eqn:hatScap}
  \end{align}
  for any $t \in [\theta_{j_0}, \theta_{j_0+1})$. Then, $0 \leq
  \Datacap{t}{\theta_{j_f}} - \hatDcap{t}{\theta_{j_f}} \leq n$ and $0
  \leq \Subcap{t}{\theta_{j_f}} - \hatScap{t}{\theta_{j_f}} \leq n$.
\end{proposition}
\begin{pf}
  Here we prove only the statements about $\Datacap{t}{\theta_{j_f}}$
  as the proof of the statements for $\Subcap{t}{\theta_{j_f}}$ are
  exactly analogous to those of $\Datacap{t}{\theta_{j_f}}$.  First of
  all notice that for any $\tau_1 < \tau_2 < \tau_3$
  \begin{equation}\label{eqn:datacap-rel}
    \Datacap{\tau_1}{\tau_3} \geq \Datacap{\tau_1}{\tau_2} + 
    \Datacap{\tau_2}{\tau_3}.
  \end{equation}
  Now, let $\Tc_0 = \theta_{j_0} + \frac{ \phi^*_{j_0} }{ R_{j_0}
  }$. Clearly, from the optimality of
  $\Datacap{\theta_{j_0}}{\theta_{j_f}}$, it follows that
  \begin{equation}\label{eqn:datacap-split}
    \Datacap{\theta_{j_0}}{\Tc_0} = n \phi^*_{j_0}, \quad 
  \Datacap{\Tc_0}{\theta_{j_f}} = n \sum_{j=j_0 + 1}^{j_f - 1} \phi^*_j.
  \end{equation}
  Thus, for the special choice of $\Tc_0$, we have the stronger
  relation $\Datacap{\theta_{j_0}}{\theta_{j_f}} =
  \Datacap{\theta_{j_0}}{\Tc_0} + \Datacap{\Tc_0}{\theta_{j_f}}$. Now,
  using~\eqref{eqn:datacap-rel} twice we get
  \begin{align*}
    \Datacap{\theta_{j_0}}{\theta_{j_f}} &\geq
    \Datacap{\theta_{j_0}}{t} + \Datacap{t}{\theta_{j_f}}
    \\
    &\geq \Datacap{\theta_{j_0}}{t} + \Datacap{t}{\Tc_0} +
    \Datacap{\Tc_0}{\theta_{j_f}} ,
  \end{align*}
  which implies
  \begin{equation*}
    \Datacap{\theta_{j_0}}{\theta_{j_f}} - \Datacap{\theta_{j_0}}{t} \geq
    \Datacap{t}{\theta_{j_f}} \geq \Datacap{t}{\Tc_0} + 
    \Datacap{\Tc_0}{\theta_{j_f}} .
  \end{equation*}
  Notice that $\Datacap{t}{\Tc_0} + \Datacap{\Tc_0}{\theta_{j_f}} =
  \hatDcap{t}{\theta_{j_f}}$. Now, we compute the difference between
  the upper and lower bounds on $\Datacap{t}{\theta_{j_f}}$
  \begin{align*}
    &\Datacap{\theta_{j_0}}{\theta_{j_f}} - \Datacap{\theta_{j_0}}{t}
    - \hatDcap{t}{\theta_{j_f}}
    \\
    &= \Datacap{\theta_{j_0}}{\Tc_0} + \Datacap{\Tc_0}{\theta_{j_f}} -
    \Datacap{\theta_{j_0}}{t} - \hatDcap{t}{\theta_{j_f}}
    \\
    &= n \left[ R_{j_0} ( \Tc_0 - \theta_{j_0} ) - \floor{ R_{j_0} (t
        - \theta_{j_0}) } - \floor{ R_{j_0} (\Tc_0 - t) } \right]
    \\
    &= n \left[ - \floor{ R_{j_0} (t - \theta_{j_0}) } - \floor{ -
        R_{j_0} (t - \theta_{j_0}) } \right] \leq n ,
  \end{align*}
  where, in arriving at the second last relation, we have used $
  \floor{ R_{j_0} (\Tc_0 - t) } = \floor{ R_{j_0} (\Tc_0 -
    \theta_{j_0}) - R_{j_0} ( t - \theta_{j_0}) }$ and the fact that
  $R_{j_0} ( \Tc_0 - \theta_{j_0} ) = \phi^*_{j_0}$ is an integer. The
  statement now follows. \qed
\end{pf}

The significance of Proposition~\ref{prop:Dcap-realtime} is that it
provides a method to reuse a previously computed solution to find a
tight sub-optimal solution to the data capacity problem in
real-time. The implication is that, if one has the computational
resources, then one may solve the full optimization problem
$\Datacap{\theta_{j_1}}{\theta_{j_2}}$ for $j_1, j_2 \in
\integersnonnegative$ and use the above result to find a tight
sub-optimal solution $\hatDcap{t}{\theta_{j_2}}$ for any $t \in [
\theta_{j_1}, \theta_{j_1 + 1} ]$.

\myclearpage
\section{Event-triggered stabilization}\label{sec:ET}

In this section, we address the problem of event-triggered control
under a time-varying channel.  Section~\ref{sec:control-no-blackouts}
address the case with no channel
blackouts. Section~\ref{sec:control-blackouts} builds on this design
and analysis to deal with the presence of channel blackouts.

\subsection{Control in the absence of channel
  blackouts}\label{sec:control-no-blackouts}

In the case of no channel blackouts, the encoder may choose to
transmit at any time and, in addition, we assume the channel rate $R$
is sufficiently high at all times (the exact technical assumption is
specified later) so that there is no need to resort to the computation
of data capacity.
For this reason, we are able to consider arbitrary (i.e., not
necessarily piecewise constant) functions $t \mapsto R(t)$. Note that,
by its discrete nature, the function $t \mapsto \pbar(t)$ is always
piecewise constant. For any $p \in \integersnonnegative$, let
\begin{equation}\label{eqn:TM}
  T_M(p) = \sigma \min \{ \Gamma_1(1,1), T, T^*(p) \} ,
\end{equation}
where $\sigma \in (0, 1)$ is a design parameter, $T$ is the parameter
chosen in~\eqref{eqn:rho_def} and $T^*$ is as defined in
Lemma~\ref{lem:Tstar-unif-lbnd-Gamma2}. As we show in the sequel, if
$T_M(p)$ is an upper bound on the communication delay when $b = np$
bits are transmitted, then it is sufficient to design an
event-triggering rule that guarantees the control objective is met.

In the presence of communication delays, we need to make sure (i) that
the control objective is not violated between a transmission and the
resulting control update and (ii) that at the control update times,
the encoding error is sufficiently small to ensure future performance.
To this end, we define
\begin{subequations}\label{eqn:Lc}
  \begin{align}
    \Lc_1(t) &\triangleq \triggerPfbar \left( T_M(\pbar(t)),
      \triggerPf(t), \epsilon(t) \right) , \label{eqn:Lc1}
    \\
    \Lc_2(t) &\triangleq \triggerChbar \left( T_M(\pbar(t)),
      \triggerPf(t), \epsilon(t), \pbar(t) ) \right) , \label{eqn:Lc2}
  \end{align}
\end{subequations}
to take care of each of these requirements.  If up to $\bbar = n\pbar$
bits are transmitted at time $t$, then $\Lc_1(t)$ provides an upper
bound on the performance-trigger function $\triggerPf$ at the
reception time which would be less than $t + T_M(\pbar(t))$, while
$\Lc_2(t)$ provides an upper bound on the channel-trigger function
$\triggerCh$ if the control is updated as soon as the packet is
received.

\begin{theorem}\longthmtitle{Event-triggered control in the absence of
    blackouts} \label{thm:no-blackout}
  Suppose $t \mapsto \pbar(t)$ is piecewise constant, as
  in~\eqref{eqn:pwc-pbar}, with a uniform lower bound $1$ (i.e., no
  blackouts) and a uniform upper bound $\pmax$. Assume that
  \begin{equation}\label{eqn:R-assump}
    R(t) \geq \frac{ p }{ T_M(p) }, \ \forall p \in 
    \until{\pbar(t)}, \ \forall t .
  \end{equation}
  Consider the system~\eqref{eqn:plant_dyn} under the feedback law $u
  = K \hat{x}$, with $t\mapsto \hat{x}(t)$ evolving according
  to~\eqref{eqn:x_hat} and the sequence $\{t_k\}_{k \in
    \integersnonnegative}$ determined recursively by
  \begin{align}\label{eqn:tk_trig1}
    t_{k+1} = \min \{ t \geq \rtil_k : \ &\Lc_1(t) \geq 1 \ \lor \
    \Lc_1(t^+) \geq 1 \ \lor \ \notag
    \\
    &\Lc_2(t) \geq 1 \ \lor \ \Lc_2(t^+) \geq 1 \} .
  \end{align}
  Let $\{r_k\}_{k \in \integersnonnegative}$ and $\{\rtil_k\}_{k \in
    \integersnonnegative}$ be given as $\rtil_0 = r_0 = t_0$ and
  $\rtil_k = r_k \leq t_k + \Delta_k$ for $k \in \integerspositive$.
  Assume the encoding scheme is such that~\eqref{eqn:coding_scheme} is
  satisfied for all $t \ge t_0$.  Further assume that $\Lc_1(t_0) \leq
  1$, $\Lc_2(t_0) \leq 1$ and that~\eqref{eq:W} holds. Let
  $\uline{p_k}$ be
  \begin{equation}\label{eqn:pk_lbound_case1}
    \uline{p_k} \! \triangleq \! \min \{ p \in \integerspositive : 
    \triggerChbar \left( \frac{ p }{ R(t_k) }, \triggerPf(t_k),
      \epsilon(t_k),  p \right) \leq 1 \}.
  \end{equation}
  Then, the following hold:
  \begin{enumerate}
  \item $\uline{p_1} \leq \pbar(t_1)$. Further for each $k \in
    \integerspositive$, if $ p_k \in \integerspositive \intersect [
    \uline{p_k}, \pbar(t_k) ]$, then $\uline{p_{k+1}} \leq \pbar(t_{k+1})$.

  \item the inter-transmission times $\{ t_{k+1} - t_k \}_{k \in
      \integerspositive}$ and inter-update times $\{ \rtil_{k+1} -
    \rtil_k \}_{k \in \integerspositive}$ have a uniform positive
    lower bound,

  \item the origin is exponentially stable for the closed-loop system,
    with $V(x(t)) \leq V_d(t_0) e^{ -\beta (t - t_0)}$ for $t \geq
    t_0$.
  \end{enumerate}
\end{theorem}
\begin{pf}
  We start by establishing two claims that we later invoke to
  establish the result.

  \emph{Claim~(a):} First, we show that for any $t \geq t_0$, if
  $\triggerPf(t) \leq 1$ and $\triggerCh(t) \leq 1$ then $\Lc_1(s) <
  1$ and $\Lc_2(s) < 1$, with $s = t$ and $s = t^+$. Indeed, if
  $\triggerPf(t) \leq 1$ and $\triggerCh(t) \leq 1$, then
  Lemma~\ref{lem:Gamma1_prop} says $\Gamma_1(\triggerPf(t),
  \epsilon(t)) \geq \min\{ \Gamma_1(1,1), T\}$. Then,
  from~\eqref{eqn:TM}, \eqref{eqn:Lc1} and from
  Lemma~\ref{lem:Gamma1_prop}(iv), we see that the claim is true for
  $\Lc_1$. Again, the conditions $\triggerPf(t) \leq 1$ and
  $\triggerCh(t) \leq 1$ along with
  Lemma~\ref{lem:Tstar-unif-lbnd-Gamma2} guarantee that for any $p \in
  \integersnonnegative$, $\Gamma_2(\triggerPf(t), \epsilon(t), p) \geq
  T^*(p)$. Thus, \eqref{eqn:TM}, \eqref{eqn:Lc2} and
  Lemma~\ref{lem:Gamma2-T_sign} imply that the claim is true for
  $\Lc_2$.

  \emph{Claim~(b):} Next, we claim that for any $k \in
  \integersnonnegative$, if $\triggerPf(\rtil_k) \leq 1$ and
  $\triggerCh(\rtil_k) \leq 1$, then $\Lc_i(t_{k+1}) \leq 1$, for $i
  \in \{1,2\}$. If the signal~$\pbar$ is constant during $[\rtil_k,
  t_{k+1}]$, the claim immediately follows from Claim~(a)
  and~\eqref{eqn:tk_trig1}. Now, let us suppose there exists $\theta
  \in [\rtil_k, t_{k+1})$ at which time $\pbar$ is discontinuous,
  i.e., $\theta \in \{\theta_j\}_{j \in \integerspositive}$ as defined
  by~\eqref{eqn:pwc-pbar}. Then, from~\eqref{eqn:tk_trig1}, it is
  clear that, for $i \in \{1,2\}$, $\Lc_i(\theta) < 1$ and
  $\Lc_i(\theta^+) < 1$. This implies that there exists an interval
  $\Ic_\theta = [\theta, \theta + \epsilon)$ such that $\Lc_i(s) < 1$
  for each $s \in \Ic_\theta$ and $i \in \{1,2\}$. Then, by continuity
  of $\Lc_i$ on each interval $(\theta_j, \theta_{j+1}]$ and by
  invoking induction over the discontinuity times of $\pbar$, we can
  conclude that the claim is true.

  Now, we show that (i) holds. The facts $\Lc_1(t_0) \leq 1$ and
  $\Lc_2(t_0) \leq 1$ together with the arguments used above ensure
  that $\Lc_1(t_1) \leq 1$ and $\Lc_2(t_1) \leq
  1$. Then,~\eqref{eqn:pk_lbound_case1} ensures that $\uline{p_1} \leq
  \pbar(t_1)$. Now, for each $k \in \integerspositive$, if $\Lc_1(t_k)
  \leq 1$ and $\Lc_2(t_k) \leq 1$ and $ p_k \in \integerspositive
  \intersect [ \uline{p_k}, \pbar(t_k) ]$ then
  \begin{equation}\label{eqn:comm-time-bnd}
    \rtil_k - t_k = r_k - t_k \leq \frac{ p_k }{ R(t_k) } 
    \leq \frac{ \pbar(t_k) }{ R(t_k) } \leq T_M(\pbar(t)) ,
  \end{equation}
  where the last inequality follows from~\eqref{eqn:R-assump}. As a
  result of~\eqref{eqn:comm-time-bnd}, we see that
  $\triggerPf(\rtil_k) \leq 1$ and $\triggerCh(\rtil_k) \leq 1$. Then,
  invoking Claim~(b), we see that $\Lc_2(t_{k+1}) \leq 1$, from which
  it follows that $\uline{p_{k+1}} \leq \pbar(t_{k+1})$, which proves
  (i).

  Now, we prove (ii) - the main idea here is that for each $k \in
  \integersnonnegative$, either $\rtil_k - t_k$ or $t_{k+1} - \rtil_k$
  is sufficiently large to guarantee (ii). To show this, we pick a
  $\sigma_1 \in (0,1)$ and partition the set $\integersnonnegative$
  into two subsets $G$ and $L$ defined as follows
  \begin{align*}
    G &= \{ k \in \integersnonnegative : \rtil_k - t_k > \sigma_1
    T_M(p_k) \} ,
    \\
    L &= \{ k \in \integersnonnegative : \rtil_k - t_k \leq \sigma_1
    T_M(p_k) \} .
  \end{align*}
  Then, it is clear that $\{ t_{k+1} - t_k \}_{k \in G}$ and $\{
  \rtil_{k+1} - \rtil_k \}_{k \in G}$ are uniformly lower bounded by
  $\sigma_1 T_M(1)$. Thus, all that remains is to handle the
  set~$L$. Recall that the assumptions and the design are such that,
  for each $k \in \integersnonnegative$, we guarantee
  $\triggerPf(\rtil_k) \leq 1$ and $\triggerCh(\rtil_k) \leq 1$ for
  $\rtil_k \leq t_k + T_M(p_k)$. As a result, and due to the fact that
  $\{p_k\}$ is upper bounded by $\pmax$, there exist $\triggerPf^0,
  \triggerCh^0 \in (0,1)$ such that $\triggerPf(\rtil_k) \leq
  \triggerPf^0$ and $\triggerCh(\rtil_k) \leq \triggerCh^0$ for all $k
  \in L$. Thus, from Claim~(a) and~\eqref{eqn:tk_trig1}, it is clear
  that for any $k \in L$, $t_{k+1} - \rtil_k \geq T_L$, where $T_L$ is
  a lower bound on the time it takes $\triggerPf$ to evolve from
  $\triggerPf^0$ to $1$ and on the time it takes $\triggerCh$ to
  evolve from $\triggerCh^0$ to $1$. Finally, by the fact that both
  $\triggerPf^0$ and $\triggerCh^0$ are strictly less than $1$, it
  follows that $T_L > 0$, which proves (ii).

  Regarding (iii), we have already seen that for any $k \in
  \integersnonnegative$, $\triggerPf(t) \leq 1$ for all $t \in [t_k,
  \rtil_k]$. Further, \eqref{eqn:tk_trig1} also ensures that
  $\triggerPf(t) \leq 1$ for all $t \in [\rtil_k, t_{k+1}]$. Therefore
  $\triggerPf(t) \leq 1$ ($V(x(t) \leq V_d(t)$) for all $t \geq t_0$,
  which completes the proof. \qed
\end{pf}

The interpretation of the three claims of the result is as
follows. Claim (i) essentially states that if the number of bits
transmitted in the past is according to the given recommendation, then
in the future, the sufficient number of bits $\uline{b_k} = n
\uline{p_k}$ to guarantee continued performance will respect the
time-varying channel constraints. Claim (ii) is sufficient to
guarantee non-Zeno behavior and claim (iii) states that indeed the
control objective is met.

\begin{remark}\longthmtitle{Requirements on the knowledge of channel
    information} {\rm Note that in the scenario with no channel
    blackouts, the encoder needs to know the channel information given
    by $R$ and $\pbar$ only over a time horizon of length
    $\delta_t$. Further, if a uniform lower bound on $t \mapsto
    \pbar(t)$ greater than or equal to $1$ is known, then it is
    sufficient for the encoder to know only the channel information at
    the current time and use this bound to schedule the transmissions
    (however, this might result in more frequent transmissions with
    smaller packet sizes).
  } \oprocend
\end{remark}

\subsection{Control in the presence of channel
  blackouts}\label{sec:control-blackouts}

Here, we address the scenario of channel blackouts building on our
developments in Section~\ref{sec:control-no-blackouts}. The main
difficulty comes from the fact that in the presence of blackouts, the
channel might be completely unavailable. Thus, the event-triggering
condition not only needs to be based on the functions $\Lc_1$ and
$\Lc_2$ in~\eqref{eqn:Lc}, but also on the available data capacity up
to the next blackout.

Throughout the section, we assume both $R$ and $\pbar$ are piecewise
constant functions, as in~\eqref{eqn:pwc-R-pbar} and, without loss of
generality, that time slots with $\pbar = 0$ are not consecutive. We
let $B_k \triangleq (\theta_{j_k}, \theta_{j_k + 1}]$ denote the
$k^{\text{th}}$ blackout slot, with $k \in \integerspositive$. Also,
for any $t \geq t_0$, we let
\begin{align*}
  \tau_l(t) & \triangleq \min \{ s \geq t : \pbar(s) = 0 \} ,
  \\
  \tau_u(t) & \triangleq \min \{ s \geq \tau_l(t) : \pbar(s) > 0 \} ,
\end{align*}
give, respectively, the beginning and the end times of the next
channel blackout slot from the current time~$t$. When there is no
confusion, we simply use $\tau_l$ and $\tau_u$, dropping the
argument~$t$. Hence, for $t \in [t_0, \theta_{j_1})$, we have
$\tau_l(t) = \theta_{j_1}$ and $\tau_u(t) = \theta_{j_1 +
  1}$. Similarly, for any $k \in \integerspositive$ and $t \in
(\theta_{j_k}, \theta_{j_{k+1}} ]$, we have $\tau_l(t) =
\theta_{j_{k+1}}$ and $\tau_u(t) = \theta_{j_{k+1}+1}$.
At time $t$, the length of the next channel blackout slot, $T_b(t)
\triangleq \tau_u (t) - \tau_l (t)$, determines a sufficient upper
bound on the encoding error $d_e(\tau_l)$, or equivalently
$\epsilon(\tau_l)$, for non-violation of the control objective during
the blackout or immediately subsequent to it. We quantify this upper
bound in the following result.

\begin{lemma}\longthmtitle{Upper bound on required $\epsilon$ before
    blackout}\label{lem:epsilon-req-blacout}
  For $t \in [t_0,\infty)$, suppose
  \begin{equation}\label{eqn:epsilonr-def}
    \epsilon(\tau_l(t)) \leq \epsilon_r(t) \triangleq \min \left\{ \frac{
        (e^{w T_b(t)} - 1 ) (w + \mu) }{ W (e^{(w+\mu) T_b(t)} - 1 ) },
      \frac{ 1 }{ e^{\bar{\mu}T_b(t)} } \right\} ,
  \end{equation}
  where $\bar{\mu} \triangleq \Infnorm{A} + \frac{\beta}{2}$. If
  $\triggerPf(\tau_l (t)) \leq 1$, then $\triggerPf(s) \leq 1$ for all
  $s \in [\tau_l(t), \tau_u(t)]$ and $\triggerCh(\tau_u(t)) \leq 1$
  (in particular $\epsilon(\tau_u(t)) \leq 1$).
\end{lemma}
\begin{pf}
  From Lemma~\ref{lem:Gamma1_prop}, we know $\Gamma_1(
  \triggerPf(\tau_l), \epsilon(\tau_l) ) \geq \Gamma_1( 1,
  \epsilon_r(t) )$. So, we need to show that $\Gamma_1 ( 1,
  \epsilon_r(t) ) \geq T_b(t)$ or, as per
  Lemma~\ref{lem:Gamma1_prop}(iv), that $\triggerPfbar(T_b(t), 1,
  \epsilon_r(t)) \leq 1$. Direct computation shows that this is indeed
  the case, which implies $\triggerPf(s) \leq 1$ for all $s \in
  [\tau_l, \tau_u]$ by the definition of~$\Gamma_1$. The second claim
  follows from
  \begin{align*}
    \triggerCh(\tau_u) &\leq \triggerChbar(T_b(t), 1, \epsilon_r(t), 0
    )
    \\
    &= \Infnorm{ e^{A T_b(t)} } e^{\frac{\beta}{2} T_b(t)}
    \epsilon_r(t) \leq e^{\bar{\mu}T_b(t)} \epsilon_r(t) \leq 1
    . \quad \square
  \end{align*}
\end{pf}

The ability to ensure that $\epsilon(\tau_l)$ is sufficiently small is
determined by the data capacity $\Datacap{t}{\tau_l}$. To have a
real-time implementation, we make use of the sub-optimal lower bound
$\hatScap{t}{\tau_l}$ instead. However, notice that maximizing the
data throughput and satisfying the primary control goal of exponential
convergence at a desired rate may not be compatible in general - if
maximizing data throughput is the only goal, then certain
transmissions might be delayed and this might lead to the violation of
the primary control objective. Conversely, if the control objective is
the only goal, this might lead to an inefficient use of the channel
that could be detrimental later.  Thus, to still be able to use the
intuition and the building blocks from
Section~\ref{sec:control-no-blackouts}, we need to impose a
time-varying artificial bound on the allowed packet size in place of
$\pbar(t)$ that prevents the system from affecting the data capacity
until the next blackout.  To this end, we store in the variable
$\Pc_j$ the value of $\phi^N_j$, where $\phi^N$ is as defined in
Section~\ref{sec:efficient-approx} for
$\Subcap{\theta_j}{\tau_l(\theta_j)}$.  Then, we define
\begin{equation}\label{eqn:Phi-def}
  \optxbits{t} \triangleq \losat{ \left \lfloor
      \Pc_j - R_j ( t - \theta_j )
    \right \rfloor }, \  t \in ( \theta_j, \theta_{j+1} ] .
\end{equation}
We notice from~\eqref{eqn:hatScap} that $n \optxbits{t}$ is the
optimal number of bits to be transmitted during $(t, \theta_{j+1} ]$
to obtain the sub-optimal data capacity $\hatScap{t}{\tau_l(t)}$. Note
that some of $n \optxbits{t}$ bits may be received after
$\theta_{j+1}$. Now, we let
\begin{equation}\label{eqn:psi-def}
  \packbnd{t} \triangleq \min \{ \pbar(t), \optxbits{t} \}
\end{equation}
be the artificial bound on the packet size for transmissions. Notice
that $\optxbits{t}$ may at times be zero, even when $\pbar(t) > 0$,
which means letting $\packbnd{t}$ be the bound on packet size may
itself introduce \emph{artificial blackouts}. However, we can state
how long artificial blackouts may be, as the next result shows.

\begin{lemma}\longthmtitle{Upper bound on the length of artificial
    blackouts}\label{lem:artif-black-bnd}
  Let $\Btil_j \triangleq \{ t \in I_j = (\theta_j, \theta_{j+1} ] :
  \packbnd{t} = 0 \}$. Then, for each $j \in \integersnonnegative$,
  $\Btil_j$ is an interval and if $\pibar_j > 0$, then the length of
  $\Btil_j$ is less than $2/R_j = 2/R(\theta_{j+1})$.
\end{lemma}
\begin{pf}
  The fact that $\Btil_j$ is an interval follows directly from the
  definition~\eqref{eqn:Phi-def}. If $\pibar_j > 0$, then at any time
  $t \in I_j$, $\pbar(t) = \pibar_j > 0$. Thus, if $\packbnd{t} = 0$
  for some $t \in I_j$,
  \begin{align*}
    &\Pc_j - R_j ( t - \theta_j ) = \Pc_j - R_j ( t + T_j -
    \theta_{j+1} ) < 1
    \\
    &\implies ( \theta_{j+1} - t ) < \frac{1}{R_j} + \left( T_j -
      \frac{ \Pc_j }{ R_j } \right) < \frac{2}{R_j} ,
  \end{align*}
  where the last inequality follows from the optimality of
  $\Subcap{\theta_j}{\tau_l(\theta_j)}$ because otherwise, if $R_j T_j
  - \Pc_j \geq 1$, then the optimality of $\Pc_j$ would imply that
  $\Pc_j = \Pc_j + 1$, which is a contradiction. This proves the
  result. \qed
\end{pf}


With this in place, we define functions analogous to $\Lc_1$ and
$\Lc_2$ to, respectively, monitor the compliance with the control
objective and ensure the encoding error is sufficiently small at the
control update times to ensure future performance.  In addition, we
define one more function to capture the effect of the data capacity,
\begin{subequations}\label{eqn:trig_funs_blackout}
  \begin{align}
    \Lctil_1(t) &\triangleq \triggerPfbar \left( \Tc(t),
      \triggerPf(t), \epsilon(t) \right) ,
    \\
    \Lctil_2(t) &\triangleq \triggerChbar \left( \Tc(t),
      \triggerPf(t), \epsilon(t), \packbnd{t} \right) ,
    \\
    \Lc_3(t,\epsilon) &\triangleq n \log_2 \left( \frac{ e^{\bar{\mu}
          (\tau_l(t) - t)} \epsilon }{ \epsilon_r(t) } \right) -
    \sigma_1 \hatScap{t}{\tau_l(t)} ,
  \end{align}
\end{subequations}
where $\sigma_1 \in (0, 1)$ is a design parameter and
\begin{equation*}
  \Tc(t) \triangleq
  \begin{cases}
    T_M(\packbnd{t}), \quad &\text{if } \packbnd{t} \geq 1
    \\
    \frac{2}{R(t)}, \quad &\text{if } \packbnd{t} = 0 .
  \end{cases}
\end{equation*}

Clearly, we cannot satisfactorily control the system for arbitrary
channel characteristics with arbitrary channel blackout slots. The
following result presents a sufficient condition on the length of the
blackout slots and the available data capacity.

\begin{lemma}\longthmtitle{Control feasibility in the presence of
    blackouts}\label{lem:con-feas-blacout}
  Suppose $t \mapsto R(t)$ and $t \mapsto \pbar(t)$ are piecewise
  constant functions as in~\eqref{eqn:pwc-R-pbar}. Let $\{
  (\theta_{j_k}, \theta_{j_k + 1}] \}_{k \in \integerspositive}$ be a
  sequence of channel blackout slots. Assume that $\pbar(t_0) > 0$,
  $\Lc_3(t_0, \epsilon(t_0)) \leq 0$ and, for each $k \in
  \integerspositive$, assume $\Lc_3(\theta_{j_k+1}, 1) \leq 0$. Then,
  there exists a transmission policy that ensures
  $\epsilon(\theta_{j_k}) \leq \epsilon_r(\theta_{j_k})$ for each $k
  \in \integerspositive$.
\end{lemma}
\begin{pf}
  Notice from the definition of $\epsilon(t)$
  in~\eqref{eqn:eps-btild-def} and~\eqref{eqn:coding_scheme} that for
  any $k \in \integersnonnegative$ and $s \in [r_k, r_{k+1})$
  \begin{equation*}
    \epsilon(s) = \frac{ \Infnorm{e^{A(s-t_k)}} e^{(\beta/2)(s-t_k)} 
      \epsilon(t_k^-) }{ 2^{p_k} }
      \leq \frac{ e^{\bar{\mu}(s-t_k)} \epsilon(t_k^-) }{ 2^{p_k} } ,
  \end{equation*}
  which when recursively used gives us
  \begin{equation*}
    \epsilon(\tau_l(t)) \leq \frac{ e^{\bar{\mu} (\tau_l(t) - t)} \epsilon(t) }{
      2^{( \Bc(t, \tau_l(t)) / n )} } ,
  \end{equation*}
  where $\Bc(t, \tau_l(t))$ is the total number of bits communicated
  (transmitted and received) during the time interval $[t,
  \tau_l(t)]$. In other words, for any $t \geq t_0$, if
  \begin{equation}\label{eqn:Bc-rel}
    \Bc(t, \tau_l(t)) \geq n \log_2 \left( \frac{
        e^{\bar{\mu} (\tau_l(t) - t)} \epsilon(t) }{ \epsilon_r(t) }
    \right)
  \end{equation}
  ensures that $\epsilon(\tau_l(t)) \leq
  \epsilon_r(\tau_l(t))$. Initially, $\Lc_3(t_0, \epsilon(t_0)) \leq
  0$ ensures that there is enough data capacity, i.e., $\Bc(t_0,
  \theta_{j_1}) \leq \hatScap{t_0}{\theta_{j_1}}$ to
  ensure~\eqref{eqn:Bc-rel}. Lemma~\ref{lem:epsilon-req-blacout}
  guarantees that for any $k \in \integerspositive$, if
  $\epsilon(\theta_{j_k}) \leq \epsilon_r(\theta_{j_k})$ then
  $\epsilon(\theta_{j_k + 1}) \leq 1$. The final claim simply follows
  from induction and the use of the fact that $\Lc_3(\theta_{j_k+1},
  1) \leq 0$ for each $k \in \integersnonnegative$. \qed
\end{pf}

Now we are ready to present our next main result.

\begin{theorem}\longthmtitle{Event-triggered control in the presence
    of blackouts} \label{thm:blackouts}
  Suppose $t \mapsto R(t)$ and $t \mapsto \pbar(t)$ satisfy the
  assumptions of Lemma~\ref{lem:con-feas-blacout}. In addition, assume
  that $\pbar$ is uniformly upper bounded by $\pmax \in
  \integerspositive$. Also, assume
  \begin{equation}\label{eqn:R-assump2}
    R(t) \geq \frac{ (p + 2) }{ T_M(p) }, \ \forall p \in \until{\pmax},
    \ \forall t .
  \end{equation}
  Consider the system~\eqref{eqn:plant_dyn} under the feedback law $u
  = K \hat{x}$, with $t\mapsto \hat{x}(t)$ evolving according
  to~\eqref{eqn:x_hat} and the sequence $\{t_k\}_{k \in
    \integersnonnegative}$ determined recursively by
  \begin{align}\label{eqn:tk_trig2}
    t_{k+1} = \min \Big\{ &t \geq \ \rtil_k : \ \packbnd{t} \geq 1 \
    \land \notag
    \\
    \Big( &\max\{\Lctil_1(t), \Lctil_1(t^+), \Lctil_2(t),
    \Lctil_2(t^+) \} \geq 1 \notag
    \\
    &\max\{ \Lctil_3(t), \Lctil_3(t^+)\} \geq 0 \Big) \Big\} ,
  \end{align}
  where $ \Lctil_3(t) \triangleq \Lc_3(t,\epsilon(t))$. Let
  $\{r_k\}_{k \in \integersnonnegative}$ be given as $\rtil_0 = r_0 =
  t_0$ and $r_k \leq t_k + \Delta_k$ for $k \in
  \integerspositive$. Let the update times $\{\rtil_k\}_{k \in
    \integersnonnegative}$ be given as $\rtil_0 = r_0$ and for $k \in
  \integerspositive$
  \begin{equation}\label{eqn:update-times}
    \rtil_k = \min \{ t \geq r_k : \packbnd{t} \geq 1 \ \lor \ 
    \pbar(t) = 0 \} .
  \end{equation}
  Assume the encoding scheme is such that~\eqref{eqn:coding_scheme} is
  satisfied for all $t \ge t_0$.  Further assume that $\Lctil_1(t_0)
  \leq 1$, $\Lctil_2(t_0) \leq 1$ and that~\eqref{eq:W} holds. Let
  $\uline{p_k}$ be given by
  \begin{equation}\label{eqn:pk_lbound_case2}
    \uline{p_k} \! \triangleq \! \min \{ p \in \integerspositive : 
    \triggerChbar \left( T_M(p), \triggerPf(t_k), \epsilon(t_k), p
    \right) \leq 1 \}.
  \end{equation}
  Then, the following hold:
  \begin{enumerate}
  \item $\uline{p_1} \leq \packbnd{t_1}$. Further for each $k \in
    \integerspositive$, if $ p_k \in \integerspositive \intersect [
    \uline{p_k}, \packbnd{t_k} ]$, then $\uline{p_{k+1}} \leq
    \packbnd{t_{k+1}}$.

  \item the inter-transmission times $\{ t_{k+1} - t_k \}_{k \in
      \integerspositive}$ and inter-update times $\{ \rtil_{k+1} -
    \rtil_k \}_{k \in \integerspositive}$ have a uniform positive
    lower bound,

  \item the origin is exponentially stable for the closed-loop system,
    with $V(x(t)) \leq V_d(t_0) e^{ -\beta (t - t_0)}$ for $t \geq
    t_0$.
  \end{enumerate}
\end{theorem}
\begin{pf}
  Notice that~\eqref{eqn:tk_trig2} ensures that for any $k \in
  \integerspositive$, $\packbnd{t_k} \geq 1$. Now, notice
  from~\eqref{eqn:update-times} that for any $k \in
  \integerspositive$, $\rtil_k > r_k$ if and only if $\packbnd{r_k} =
  0$ and $\pbar(r_k) \geq 1$. That is, $\rtil_k > r_k$ if and only if
  $r_k \in (\tau_1, \tau_2]$, an artificial blackout interval. In all
  other cases, $\rtil_k = r_k$. Thus, it follows from
  Lemma~\ref{lem:artif-black-bnd} that $\rtil_k - r_k \leq
  \tfrac{2}{R(r_k)}$ for all $k \in \integerspositive$. Hence, for all
  $k \in \integerspositive$, we have
  \begin{align*}
    &\rtil_k - t_k = ( \rtil_k - r_k ) + ( r_k - t_k ) \leq \frac{ 2
    }{ R(r_k) } + \frac{ p_k }{ R(t_k) }
    \\
    &\implies \rtil_k - t_k \leq
    \begin{cases}
      \frac{ p_k }{ R(t_k) } , \ \text{if } \rtil_k = r_k
      \\
      \frac{ (p_k + 2) }{ \min \{ R(t_k), R(r_k) \} } , \ \text{if }
      \rtil_k > r_k .
    \end{cases}
  \end{align*}
  In either case, it follows from~\eqref{eqn:R-assump2} that $\rtil_k
  - t_k \leq T_M(p_k) \leq T_M(\packbnd{t_k})$ for all $k \in
  \integerspositive$. Thus, claims (a) and (b) in the proof of
  Theorem~\ref{thm:no-blackout} hold here also.

  Next observe that, by the construction of $t \mapsto \packbnd{t}$
  in~\eqref{eqn:psi-def}, we have $ \hatScap{\rtil_k}{\tau_l} \geq
  \hatScap{t_k}{\tau_l} - np_k$.  Next, noting that
  \begin{equation*}
    \epsilon(\rtil_k) = \Infnorm{ e^{A \Deltil_k} }
    e^{ \frac{ \beta }{ 2 } \Deltil_k } \frac{ \epsilon(t_k) }{
      2^{p_k} } \leq e^{ \bar{\mu} \Deltil_k } \frac{ \epsilon(t_k) }{
      2^{p_k} } ,
  \end{equation*}
  we have
  \begin{align*}
    &n \log_2 \left( \frac{ e^{\bar{\mu} (\tau_l - \rtil_k)}
        \epsilon(\rtil_k) }{ \epsilon_r } \right) \leq n \log_2
    \left( \frac{ e^{\bar{\mu} (\tau_l - t_k)} \epsilon(t_k) }{
        \epsilon_r } \right) - np_k
    \\
    &\leq \sigma_1 \hatScap{t_k}{\tau_l} - n p_k
    \leq \sigma_1 ( \hatScap{t_k}{\tau_l} - n p_k )
    \\
    &\leq \sigma_1 \hatScap{\rtil_k}{\tau_l} ,
  \end{align*}
  where the second inequality follows from $\Lctil_3(t_k) \leq 0$ and
  the third inequality follows from $\sigma_1 \in (0, 1)$. Therefore,
  $\Lctil_3(\rtil_k) \leq 0$. Thus, using induction, the proposed
  transmission policy ensures that by the beginning of the next
  blackout, $t = \tau_l$, $\epsilon(\tau_l) \leq
  \epsilon_r$. Lemma~\ref{lem:epsilon-req-blacout} then implies that,
  at the end of blackout, we have $\triggerCh(\tau_u) \leq 1$ and
  $\triggerPf(s) \leq 1$ for all $s \in [\tau_l, \tau_u]$. Hence,
  claim (i) follows as in the proof of
  Theorem~\ref{thm:no-blackout}(i) and using induction over the
  sequence of blackout slots.

  Claim (ii) also follows by arguments analogous to the proof of
  Theorem~\ref{thm:no-blackout}(ii).

  Finally, we prove (iii). Notice~\eqref{eqn:tk_trig2} ensures that
  $\Lctil_1(t_k) \leq 1$ for any $k \in \integerspositive$, which as a
  consequence of Lemma~\ref{lem:Gamma1_prop}(iv) means that
  $\triggerPf(t) \leq 1$ for all $t \in [t_k, \rtil_k]$ for any $k \in
  \integerspositive$. Now, for $t \in [\rtil_k, t_{k+1})$ for $k \in
  \integersnonnegative$, there are three cases. \emph{Case~I:}
  $\packbnd{t} \geq 1$. In this case, $\triggerPf(t) \leq 1$ because
  $\Lctil_1(t) < 1$. \emph{Case~II:} $\packbnd{t} = 0$ and $\pbar(t)
  \geq 1$, which corresponds to a time during an artificial blackout
  $(\tau_1, \tau_2]$. Recall from Lemma~\ref{lem:artif-black-bnd} that
  $\tau_2 - \tau_1 \leq 2/R(\tau_1)$, which
  using~\eqref{eqn:R-assump2} then implies $\tau_2 - \tau_1 \leq
  T_M(\packbnd{\tau_1^-})$. Next, by design~\eqref{eqn:update-times},
  $\rtil_k \notin ( \tau_1, \tau_2 ]$ and hence $\rtil_k < \tau_1$ and
  no transmission is in progress during $(\tau_1, \tau_2]$, which must
  mean $\Lctil_1(\tau_1^-) < 1$. Lemma~\ref{lem:Gamma1_prop}(iv) then
  implies $\Gamma_1(\triggerPf(\tau_1), \epsilon(\tau_1)) \geq
  T_M(\packbnd{\tau_1^-} \geq \tau_2 - \tau_1$. Therefore,
  $\triggerPf(t) \leq 1$ for all $t \in [\tau_1, \tau_2)$.
  \emph{Case~III:} $\packbnd{t} = \pbar(t) = 0$, which corresponds to
  a time in a channel blackout slot. We have already seen in the proof
  of (i) that the proposed transmission policy ensures $\triggerPf(s)
  \leq 1$ for all $s \in [\tau_l, \tau_u]$ for any channel black out
  $[\tau_l, \tau_u]$. Therefore, $\triggerPf(t) \leq 1$ ($V(x(t)) \leq
  V_d(t)$) for $t \geq t_0$. \qed
\end{pf}

Claim (i) in the result may be interpreted as the satisfaction of the
constraints imposed by the channel. The use of~$\psi^{\tau_l}$
in~\eqref{eqn:tk_trig2} and~\eqref{eqn:update-times} also ensures that
the data capacity is not lowered at any time in the future due to past
transmissions. The interpretation of claims (ii) and (iii) is the same
as in Theorem~\ref{thm:no-blackout}.

\begin{remark}\longthmtitle{Requirements on the knowledge of channel
    information}
  {\rm In the scenario with channel blackouts, the encoder needs to
    know, at $t \in [t_0,\infty)$, the time at which the next blackout
    will occur $\tau_l(t)$ and its duration $T_b(t)$, from which
    $\epsilon_r(t)$ may be computed. The encoder also needs to know
    the channel functions $s \mapsto R(s)$ and $s \mapsto \pbar(s)$
    for all $s \in [t, \tau_l(t)]$.  Using this information, the
    encoder can compute the lower bound on the remaining data capacity
    by computing $\hatScap{t}{\tau_l(t)}$. } \oprocend
\end{remark}

\section{Simulation results}\label{sec:sim}

\begin{figure}[!htpb]
  \centering \subfigure[\label{fig:bit_profile}]{\includegraphics[
    width=0.23\textwidth,height=1.2 in]{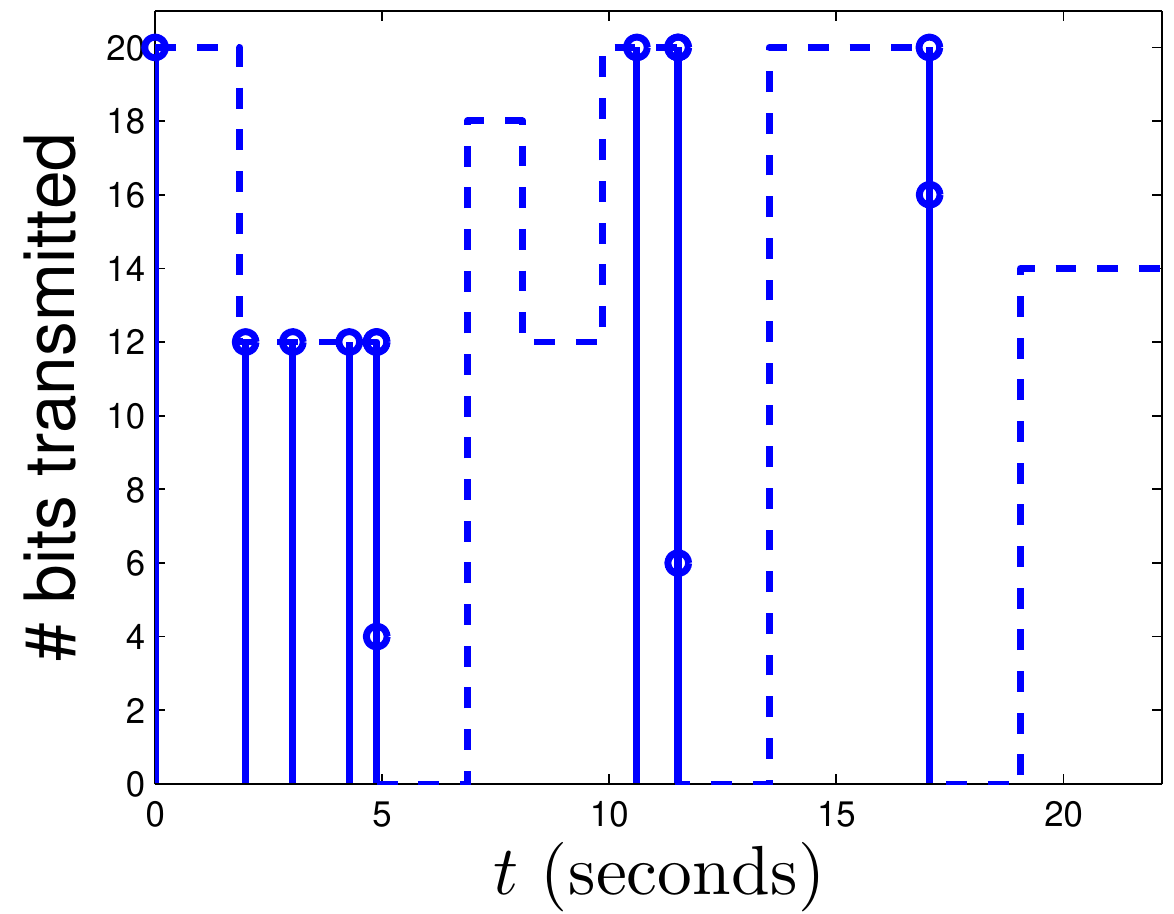}}
  \subfigure[\label{fig:R}]{\includegraphics[
    width=0.23\textwidth,height=1.3 in]{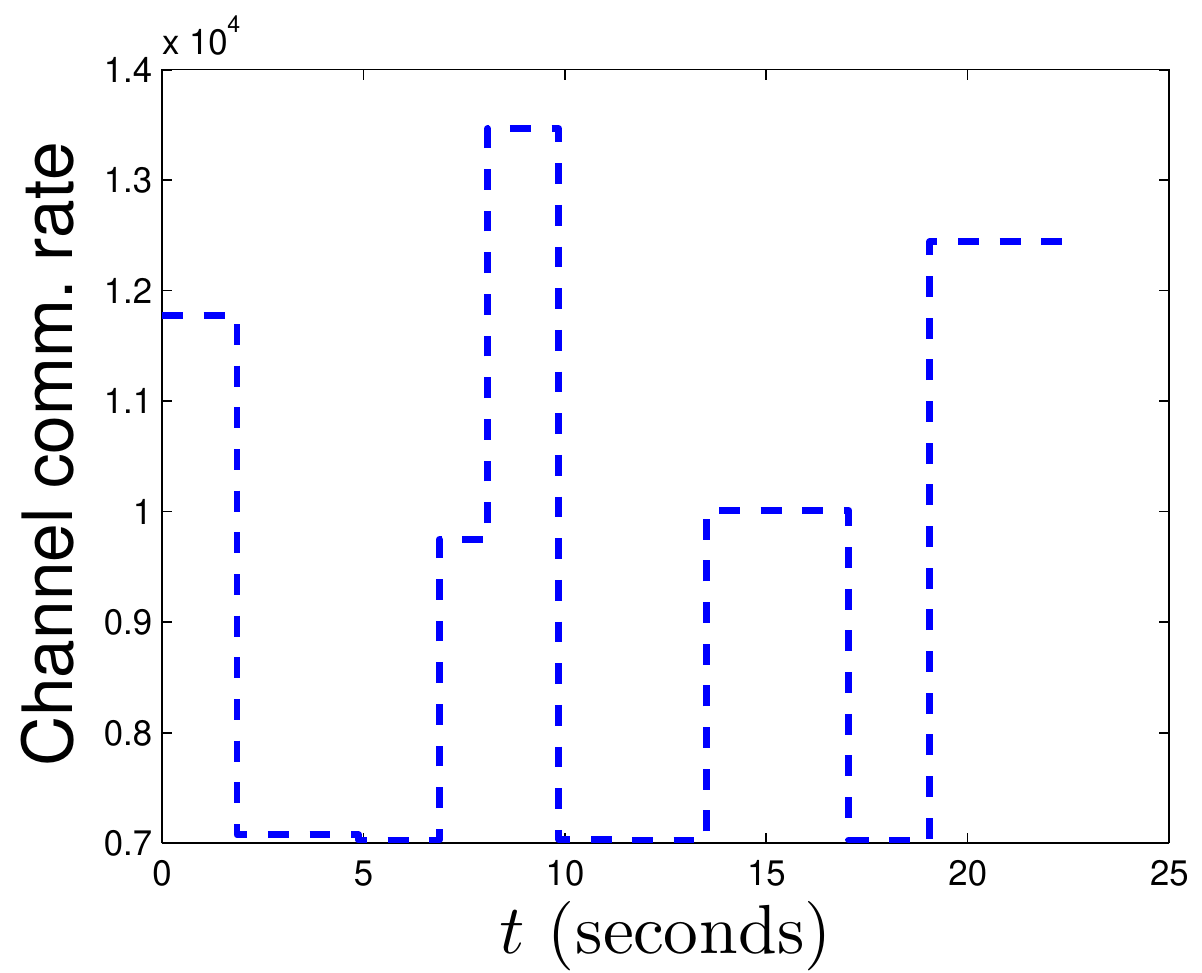}}
  \caption{(a) shows the transmission times, the number of bits
    transmitted on each transmission and the time-varying function $n
    \pbar$ (dashed line). The three intervals, $(4.88, 6.88]$,
    $(11.52, 13.52]$ and $(17.05, 19.05]$, with $\pbar = 0$ are the
    blackouts. (b) shows the time-varying function $R$. }
\end{figure}

\begin{figure}[!htpb]
  \centering \subfigure[\label{fig:V_Vd}]{\includegraphics[
    width=0.23\textwidth,height=1.2 in]{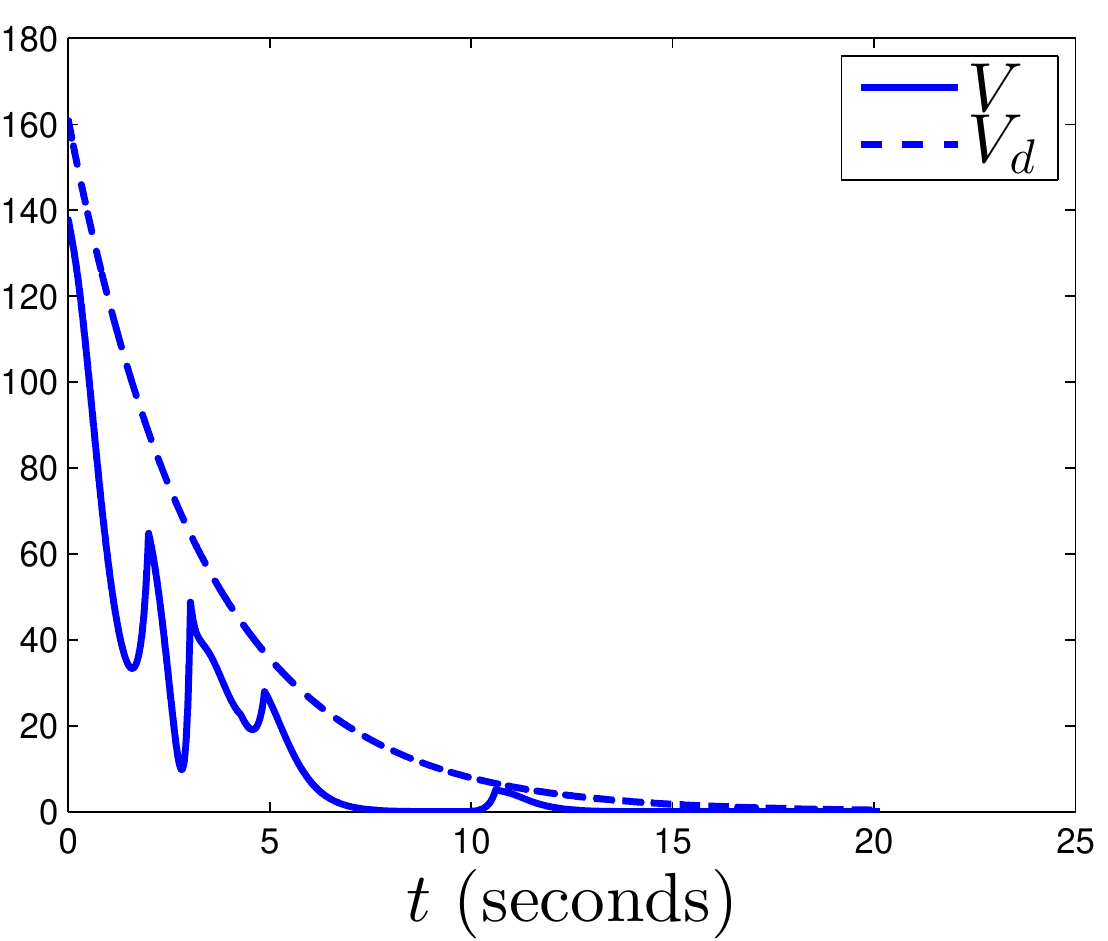}}
  \subfigure[\label{fig:cum_bit_inset}]{\includegraphics[
    width=0.23\textwidth,height=1.2 in]{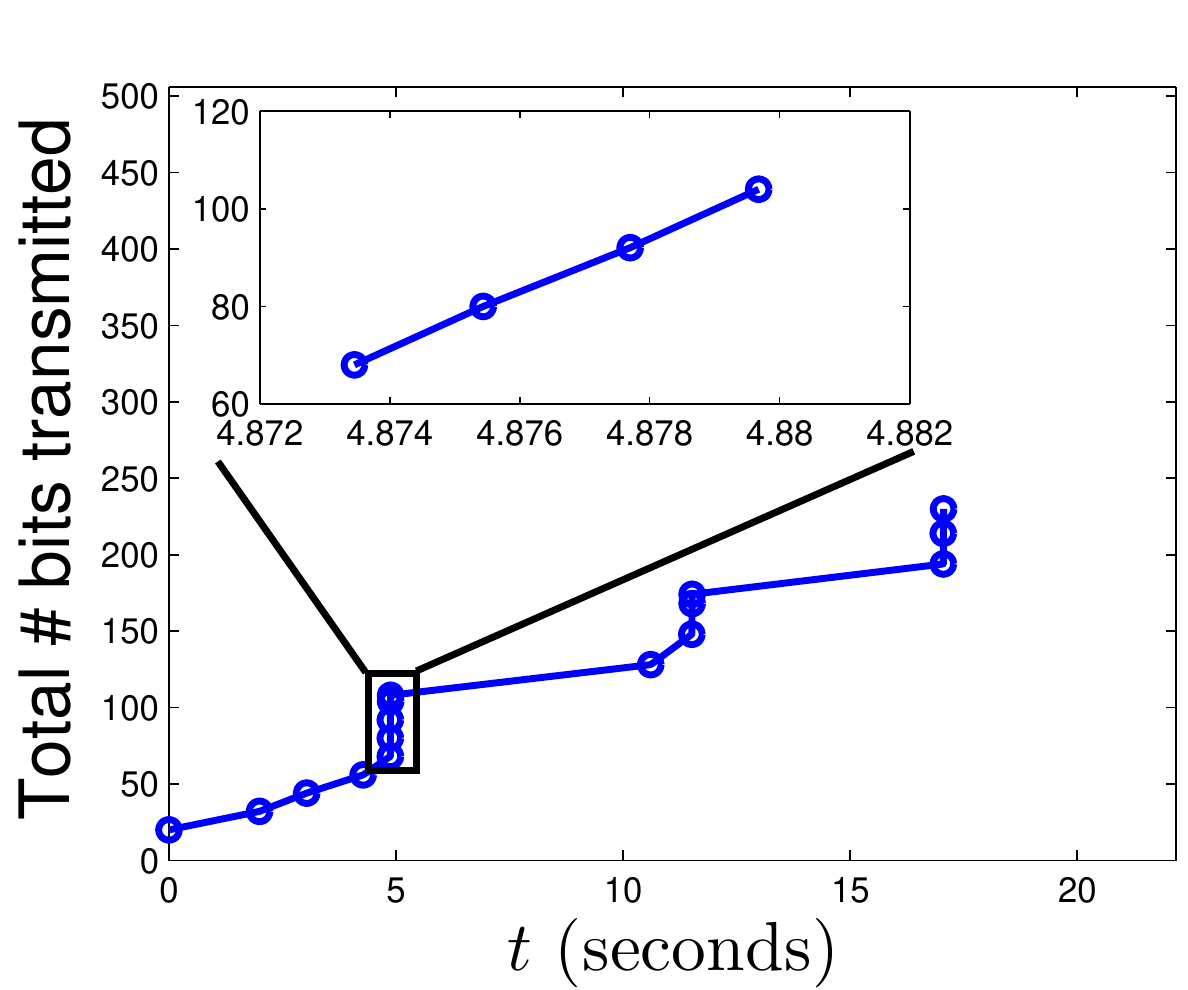}}
  \vspace*{-1ex}
  \caption{Evolution of (a) $V$ and $V_d$ and (b) total number of bits
    transmitted, the inset shows that the transmission times are
    separated. }
\end{figure}

In this section we illustrate the execution of our event-triggered
design of Section~\ref{sec:ET}.  The simulation results we present
correspond to the strategy described in Theorem~\ref{thm:blackouts} on
the system given by~\eqref{eqn:plant_dyn} with
\begin{equation*}
  A = 
  \begin{bmatrix}
    1 & -2
    \\
    1 & 4
  \end{bmatrix}, \ 
  B = 
  \begin{bmatrix}
    0
    \\
    1
  \end{bmatrix}, \ K =
  \begin{bmatrix}
    2 & -8
  \end{bmatrix} .
\end{equation*}
The plant matrix $A$ has eigenvalues at $2$ and $3$, while the control
gain matrix $K$ places the eigenvalues of the matrix $\bar{A} = A+BK$
at $-1$ and $-2$. We select the matrix $Q=\identity{2}$, for which the
solution to the Lyapunov equation~\eqref{eq:Lyap-eq} is
\begin{align*}
  P =
  \begin{bmatrix}
    2.2500 & -0.9167
    \\
    -0.9167 & 0.5833
  \end{bmatrix}
  .
\end{align*}
The desired control performance is specified by
\begin{align*}
  V_d(t_0) = 1.2 V(x(t_0)), \quad \beta = 0.8
  \frac{\lambda_m(Q)}{\lambda_M(P)} .
\end{align*}
We set $a = 1.2$ in~\eqref{eq:W}, so that $W > 0$, and assume, without
loss of generality, $t_0 = 0$.  The initial condition is $x(t_0) = (
6, -4)$, and the encoder and decoder use the information
\begin{align*}
  \hat{x}(t_0) = ( 0, 0) , \quad d_e(t_0) = 1.5 \Infnorm{x(t_0) -
    \hat{x}(t_0)} .
\end{align*}
In~\eqref{eqn:trig_funs_blackout}, we chose $\sigma_1 = 0.8$. For
these parameters, $\Gamma_1(1,1) = 0.5699$. We select $T = 0.1 \times
\Gamma(1,1)$ and $T_M(p) = 0.06 \times \min \{ \Gamma(1,1), T, T^*(p)
\}$. The time-varying channel functions $n \pbar$ and $R$ are plotted
in Figures~\ref{fig:bit_profile} and~\ref{fig:R} respectively with
dashed lines. Figure~\ref{fig:bit_profile} also shows the times of
transmission and the number of bits transmitted on each one. Note
that, in this simulation, the maximum possible number of bits are
transmitted on each transmission.
Figure~\ref{fig:V_Vd} shows the evolution of $V$ and $V_d$ and it is
clear that the control goal is satisfied. Notice that, just before a
blackout, $V$ decreases sharply in anticipation to ensure that the
control goal is not violated during the
blackout. Figure~\ref{fig:cum_bit_inset} shows the (interpolated)
cumulative number of bits transmitted as a function of time. We see
that there is a rush of transmissions just prior to $4.88$ units of
time, which we see from Figure~\ref{fig:bit_profile} is the beginning
of the first blackout.
The number of transmissions in the $20$ units of time in the
simulation are $16$, with the average inter-transmission interval as
$1.26$ and the minimum as $0.002$. From
Figure~\ref{fig:cum_bit_inset}, we also see that on an average $11.5$
bits are transmitted per unit time.

\section{Conclusions}\label{sec:conc}

We have addressed the problem of event-triggered control of linear
time-invariant systems under time-varying rate-limited communication
channels. The class of time-varying channels we consider is broad
enough to include intermittent occurrence of channel blackouts, which
are intervals of time when the communication channel is unavailable
for feedback. We have designed an event-triggered control scheme that,
using prior knowledge of the channel information, guarantees the
exponential stabilization of the system at a desired convergence rate,
even in the presence of intermittent channel blackouts.  Key enablers
of our design are the definition and analysis of the data capacity,
which measures the maximum number of bits that can be communicated
over a given time interval through one or more transmissions. We have
also provided an efficient real-time algorithm to lower bound the data
capacity for a time-slotted model of channel evolution. An important
assumption we make is that the encoder has knowledge of the channel
evolution sufficiently ahead of time so that it can plan its
transmission schedule. In practice, the channel will have to be
estimated, and only uncertain knowledge of its future evolution may be
available. Nevertheless, we showed that the problem of estimating the
data capacity, which is needed in order to design a meaningful
mechanism to guarantee exponential stability, is challenging even
assuming full channel information. Future work will explore the
reduction of the conservatism of the proposed design, scenarios with
bounded disturbances, a stochastic model of channel evolution, and the
trade-off between the available information pattern at the encoder and
the ability to perform event-triggered control.


\bibliographystyle{plainnat}
\bibliography{alias,FB,JC,Main,Main-add}

\end{document}